\documentclass[prd,aps,showpacs,epsf,floats,onecolumn]{revtex4}%
\usepackage{amssymb}
\usepackage{amsfonts}
\usepackage{amsmath}
\usepackage{graphicx}
\usepackage{fancyhdr}%
\providecommand{\U}[1]{\protect\rule{.1in}{.1in}}
\begin{document}
\title{\textbf{Optimal-Speed Unitary Quantum Time Evolutions and Propagation of Light
with Maximal Degree of Coherence}}
\author{\textbf{Carlo Cafaro}$^{1}$, \textbf{Shannon Ray}$^{2}$, and \textbf{Paul
M.\ Alsing}$^{2}$}
\affiliation{$^{1}$SUNY Polytechnic Institute, Albany, New York 12203, USA}
\affiliation{$^{2}$Air Force Research Laboratory, Information Directorate, Rome, New York
13441, USA}

\begin{abstract}
It is recognized that Grover arrived at his original quantum search algorithm
inspired by his comprehension of the interference of classical waves
originating from an array of antennas. It is also known that
quantum-mechanical characterization of electromagnetic radiation is isomorphic
to the treatment of the orientation of a spin-$1/2$ particle. In this paper,
motivated by Grover's original intuition and starting from this mathematical
equivalence, we present a quantitative link\textbf{ }between the geometry of
time-independent optimal-speed Hamiltonian evolutions on the Bloch sphere and
the geometry of intensity-preserving propagation of light with maximal degree
of coherence on the Poincar\'{e} sphere. Finally, identifying interference as
the fundamental physical ingredient underlying both physical phenomena, we
propose that our work can provide in retrospect a quantitative geometric
background underlying Grover's powerful intuition.

\end{abstract}

\maketitle

\fancyhead[R]{\ifnum\value{page}<2\relax\else\thepage\fi}

\thispagestyle{fancy}

\section{Introduction}

One of the main goals of quantum information science (QIS), including quantum
nanoengineering and quantum optics, is the development of devices capable of
reliably processing quantum information \cite{caneva09,carlini13,heger13}.
When considering the implementation of such quantum technologies, it becomes
especially relevant engineering a suitable Hamiltonian that evolves an initial
source state into a final target state. An essential condition for these
quantum information processing devices is the capacity of having total control
on the state of a single qubit on time scales much shorter than the coherence
time. For such reasons, the conceptual understanding of controlled quantum
dynamics along with their limits is becoming increasingly important in QIS.
The cost functional that quantifies the efficiency of getting to the target
state from a given initial state depends on the physical scenario being
considered. In the simplest case, one can focus on an unconstrained
Hamiltonian time-evolution, except for a bound on the energy resource. Then,
regarding time-optimality as the cost functional, the problem becomes finding
the time-independent Hamiltonian that generates maximum speed of evolution.
However, there can be a range of constraints that forbids the implementation
of such an elementary protocol in more realistic scenarios. Indeed, in real
laboratory settings needed for the implementation of quantum technologies, one
would need to apply various optimization techniques available within the more
general framework of optimal quantum control theory \cite{boscain21} to
identify suitable time-dependent Hamiltonians that generate the dynamics
achieving required quantum tasks. The transition to time-dependent
Hamiltonians can be motivated by several reasons, including the presence of
time-varying external magnetic fields \cite{brody15} or, alternatively, the
existence of dissipation due to a coupling between the quantum system and the
environment \cite{deffner14}. From this wide range of physical scenarios that
one could take into consideration, we shall focus in this paper on the
simplest case, namely that of time-optimal quantum mechanical unitary
evolution in the presence of a bound on the energy resource.

It is known that the quantum-mechanical treatment of photon polarization is
mathematically equivalent to the treatment of the orientation of a spin-$1/2$
particle \cite{fano54}. In particular, focusing on the physics of two-level
quantum systems and classical polarization optics in two-dimensions, the
concepts of Bloch vector and Bloch sphere \cite{nielsen} are the analogs of
the notions of Stokes vector and Poincar\'{e} sphere \cite{born50}, respectively.

A remarkable link between quantum mechanics and classical optics is
represented by the interpretation of Pancharatnam's optical phase that appears
in the context of interference of polarized light \cite{pancharatnam56} as an
early example of Berry's (nondynamical) geometric phase that emerges in the
context of cyclic and adiabatic quantum mechanical evolutions \cite{berry84}.
For an in-depth discussion on the relation between Pancharatnam's phase and
Berry's phase, we refer to Refs. \cite{berry87,samuel88}. The work in Ref.
\cite{berry87} is especially illuminating since Berry, starting from the
description of polarization in terms of the Poincar\'{e} sphere, expresses
Pancharatnam's classical optics analysis in quantum mechanical language and,
moreover, clarifies the relation between the classical optical phase and the
quantum adiabatic phase. Interestingly, this mutual interaction between
quantum mechanics and classical optics has been rather beneficial in science.
For example, borrowing ideas from Pancharatnam's work on the classical
interference of polarized light, Samuel and Bhandari extended the concept of
Berry's phase to nonunitary and noncyclic quantum mechanical evolutions in
Ref. \cite{samuel88}. Furthermore, just as Pancharatnam's theory was tested in
an experimental fashion with the detection of the predicted phase shifts by
interference, the first experimental manifestation of Berry's phase was
carried out in an optical experiment \cite{wu86} where the Berry phase
measured corresponded to an angle of rotation of a plane of polarization of
light \cite{jordan87}. A second close similarity between quantum mechanics and
classical optics is the correspondence between the degree of polarization of
beams of light and the parity of qubits as reported in Refs.
\cite{cessa08,james12}. The origin of this similarity can be explained as
follows. In the quantum mechanical Bloch sphere formalism, the origin
represents a maximally mixed state, whereas points on the surface of the
sphere are pure states. In the Poincar\'{e} sphere formalism in classical
optics, the origin represents a completely unpolarized light beam, whereas
points on the surface of the sphere are completely polarized beams.

In quantum mechanics, there are several ways in which one can derive an
expression of time-independent optimal-speed Hamiltonians evolving an initial
state $\left\vert A\right\rangle $ into a final state $\left\vert
B\right\rangle $. For instance, a simple derivation can rely on finding
Hamiltonians $\left\{  \mathrm{H}\right\}  $ that evolve $\left\vert
A\right\rangle $ into $\left\vert B\right\rangle $ in the least time subject
to the constraint that the difference between the largest and the smallest
eigenvalues of $\mathrm{H}$ is held fixed \cite{bender07}. Another
straightforward derivation, instead, can put the emphasis on choosing the
Hamiltonian so that the uncertainty in energy is maximized \cite{ali09}. In
any case, the trajectories connecting $\left\vert A\right\rangle $ and
$\left\vert B\right\rangle $ generated by such optimal-speed unitary
evolutions $\left\{  U\right\}  $ can be viewed as geodesic curves on the
Bloch sphere (or, alternatively, the two-sphere $S^{2}$). For this reason, it
is especially interesting the geometric interpretation of these unitary
operators $\left\{  U\right\}  $ with $\left\vert A\right\rangle $
$\overset{U}{\rightarrow}\left\vert B\right\rangle $ in terms of rotations of
the Bloch sphere around the axis that is orthogonal to the hemispherical plane
containing the origin along with $\left\vert A\right\rangle $ and $\left\vert
B\right\rangle $. In particular, the Hamiltonian that generates the rotation
takes the form $\mathrm{H}=E_{+}\left\vert E_{+}\right\rangle \left\langle
E_{+}\right\vert +E_{-}\left\vert E_{-}\right\rangle \left\langle
E_{-}\right\vert $ for a pair of real parameters $E_{\pm}$ with the axis of
rotation corresponding to a pair of orthogonal states $\left\vert E_{\pm
}\right\rangle $ \cite{brody06,brody07}. More generally, given that the
dynamics induced by a optimal-speed unitary evolution can be regarded as a
rigid rotation of the two-sphere $S^{2}$, if there exists a unitary evolution
transforming $\left\vert A\right\rangle $ into $\left\vert B\right\rangle $
with $\left\langle A|B\right\rangle =0$ along a geodesic path, then there must
exist a pair of energy eigenstates $\left\vert E_{+}\right\rangle $ and
$\left\vert E_{-}\right\rangle $, say at the equator of $S^{2}$, such that
$\left\vert A\right\rangle $ and $\left\vert B\right\rangle $ lie at the poles
of $S^{2}$ \cite{brody03}. Moreover, in terms of an efficiency measure defined
by means of the ratio between the distance along the shortest geodesic path
joining $\left\vert A\right\rangle $ and $\left\vert B\right\rangle $ and the
distance along the actual dynamical trajectory traced by the state vector
$\left\vert \psi\left(  t\right)  \right\rangle \overset{\text{def}}%
{=}e^{-\frac{i}{\hslash}\mathrm{H}t}\left\vert A\right\rangle $, these
optimal-speed unitary quantum mechanical evolutions exhibit unit
\textquotedblleft\emph{quantum geometric efficiency}\textquotedblright%
\ \cite{anandan90,cafaro20A}.

In classical polarization optics, it is known that the degree of polarization
of a light wave propagating along the $\hat{z}$-direction does not depend on
the choice of the $\hat{x}$- and $\hat{y}$-directions. Furthermore, such a
degree of polarization is an upper bound for the so-called degree of coherence
between the electric vibrations in the $\hat{x}$- and $\hat{y}$-directions
\cite{born50}. Interestingly, it can be demonstrated that there always exists
a pair of orthogonal directions for which the degree of coherence has its
maximum value and this value is equal to the degree of polarization of the
light wave \cite{wolf59}. Therefore, considering the ratio between the degree
of coherence and the degree of polarization as some sort of \textquotedblleft%
\emph{classical optical efficiency}\textquotedblright, it happens that there
is always an optimal optical configuration in which the propagation of
polarized light occurs with maximal degree of coherence.

In this paper, we wish to investigate an unexplored link between optimal-speed
quantum mechanical evolutions and propagation of light with maximal degree of
coherence. Our investigation is inspired by the above mentioned existing links
between quantum mechanics and classical optics. Furthermore, we rely on our
familiarity with both digital and analog quantum search algorithms
\cite{cafaro17,cafaro19A,cafaro19B,cafaro20A,gassner20,cafaro20B}. In
addition, our proposed investigation finds additional motivation by recalling
that Grover's original intuition that helped him creating his quantum search
algorithm \cite{grover97} was based upon a classical optics
phenomenon.\ Specifically, Grover arrived at his quantum search algorithm by
observing the interference of classical waves originating from an array of
antennas \cite{lloyd99}. In this way, by mimicking the interference of
\emph{classical} waves, Grover arrived at his\emph{ quantum} search scheme.

Therefore, motivated by this intriguing similarity between the existence of a
convenient pair of orthogonal energy eigenstates in the geometrical
description of optimal-speed quantum evolutions and the existence of a
suitable pair of orthogonal directions for the electric field in the geometric
description of propagation of polarized light with optimal-coherence, we
provide in this paper a quantitative link between quantum mechanics and
classical polarization optics. Specifically, starting from the mathematical
equivalence between the quantum-mechanical characterization of electromagnetic
radiation and the treatment of the orientation of a spin-$1/2$ particle, we
discuss in a quantitative manner the connection between the geometry of
time-independent optimal-speed Hamiltonian evolutions on the Bloch sphere and
the geometry of intensity-preserving propagation of light with maximal degree
of coherence on the Poincar\'{e} sphere. Identifying interference as the
essential physical ingredient underlying both phenomena \ being studied in our
paper, we conclude by arguing that our work can provide a quantitative
geometric background underlying Grover's powerful intuition.

To summarize, we are aware of the following known links: (1) Connection
between interference in classical wave theory and interference in quantum
mechanics and (2) equivalence between the Bloch sphere and the Poincar\'{e}
sphere. In this paper, however, we provide a link between the geometry of the
unitary dynamics of time-independent optimal-speed Hamiltonians on the Bloch
sphere and the geometry of intensity-preserving propagation of light with
maximal degree of coherence on the Poincar\'{e} sphere. Our finding
establishes a bridge between the physics of two-level quantum systems and the
physics of classical polarization optics via the discovery of this previously
unknown connection.

The layout of the remainder of this paper is as follows. In Sec. II, we
discuss two alternative characterizations of optimal-speed Hamiltonians. The
first analysis focuses on minimizing the evolution time subject to the energy
eigenvalue constraint. The second one, instead, relies on the maximization of
the energy uncertainty that yields the spectral decomposition of the optimal
Hamiltonian. We conclude Sec. II with a discussion on unit geometric
efficiency on the Bloch sphere. In Sec. III, after providing some motivational
background, we characterize the propagation of light in terms of the
polarization ellipse, the Stokes parameters, and the Poincar\'{e} sphere.
Then, we briefly present the concepts of coherence of electric vibrations
along with degree of coherence, coherency matrix, and degree of polarization
of a light wave. In Sec. IV, we discuss the propagation of polarized light
with maximal degree of coherence, that is propagation of light with unit
optical efficiency on the Poincar\'{e} sphere. The quantitiative link between
the geometry of time-independent optimal-speed Hamiltonian evolutions on the
Bloch sphere and the geometry of intensity-preserving propagation of light
with maximal degree of coherence on the Poincar\'{e} sphere is carried out
throughout Secs. II and IV. In Sec. V, we discuss the physical origin of our
proposed link. Our concluding remarks appear in Section VI. A number of
technical details and remarks appear in Appendixes A, B, C, and D.
Specifically, in Appendix A we place some specific properties of the Mueller
matrices in optics. In Appendix B, we employ the Poincar\'{e} sphere formalism
to describe the dependence of the modulus of the complex degree of coherence
of a partially polarized light beam in terms of the ellipticity and
orientation angles. In Appendix C, we report some mathematical details on the
parametrization of qubits and polarization states regarded as points on the
Bloch sphere and the Poincar\'{e} sphere, respectively. Finally, we present in
Appendix D a discussion on the role played by interference effects in light
propagation, quantum searching, and optimal-speed quantum evolutions.

\section{Quantum evolutions with unit geometric efficiency}

In this section, we discuss two alternative descriptions of optimal-speed
Hamiltonians. The first characterization focuses on minimizing the evolution
time subject to the energy eigenvalue constraint. The second one, instead,
depends on the maximization of the energy uncertainty that leads to the
spectral decomposition of the optimal Hamiltonian. We conclude this section
with a discussion on unit geometric efficiency on the Bloch sphere.

When studying the geometric characterization of unit efficiency
\cite{anandan90} quantum mechanical unitary evolutions specified by
time-independent Hamiltonians $\left\{  \mathrm{H}\right\}  $ under which a
normalized initial state vector $\left\vert A\right\rangle $ evolves into a
normalized final state vector $\left\vert B\right\rangle $, one notices at
least two alternative approaches in the literature. In a first approach,
researchers aim to find an expression of the Hamiltonian by minimizing the
evolution time $\Delta t\overset{\text{def}}{=}T_{AB}$ needed for evolving
$\left\vert A\right\rangle $ into $\left\vert B\right\rangle $ subject to the
constraint that the difference between the largest ($E_{+}$) and smallest
($E_{-}$) eigenvalues of the Hamiltonian is kept fixed \cite{bender07},
$E_{+}-E_{-}\overset{\text{def}}{=}E_{0}=\mathrm{fixed}$. In a second
approach, instead, investigators seek for an expression of the Hamiltonian by
maximizing the uncertainty in energy $\Delta E$ of the system \cite{ali09}.
This approach is motivated by the fact that the (angular) speed of the
minimal-time evolution $v$ of the quantum system is proportional to $\Delta
E$, $v\overset{\text{def}}{=}ds_{\mathrm{FS}}/dt\propto$ $\Delta E$, with
$s_{\mathrm{FS}}$ denoting the Fubini-Study distance between the two points on
the projective Hilbert space $\mathcal{P}\left(  \mathcal{H}\right)  $ that
corresponds to the selected initial and final states $\left\vert
A\right\rangle $ and $\left\vert B\right\rangle $, respectively. Despite the
fact that these two quantum approaches are essentially equivalent since the
constraint on the difference between the largest and the smallest eigenvalues
of the Hamiltonian is similar to upper bounding the energy uncertainty $\Delta
E$ since $\Delta E_{\max}=\left(  E_{+}-E_{-}\right)  /2$, they do put the
emphasis on distinct features that will help us better understanding the
details of the optimal evolution Hamiltonian. Ultimately, these complementary
features will help us describing the formal analogies between the geometry of
quantum evolutions with unit quantum geometric efficiency and the geometry of
classical polarization optics for light waves with degree of
polarization\textbf{ }\textrm{P}\textbf{ }that equals the degree of coherence
$\left\vert j_{xy}\right\vert $ between the electric vibrations in any two
mutually orthogonal directions of propagation of the wave \cite{wolf59}. Unit
quantum geometric efficiency means here that\textbf{ }$\eta_{\mathrm{QM}%
}\overset{\text{def}}{=}s_{0}/s=1$,\textbf{ }where $s_{0}$ is the distance
along the shortest geodesic joining the initial and final points of the
evolution that are distinct on the projective Hilbert space while $s$ is the
distance along the effective evolution of the system in the projective Hilbert
space as measured by the Fubini-Study metric. Finally, unit classical optical
efficiency means here\textbf{ }$\eta_{\mathrm{optics}}\overset{\text{def}}{=}%
$\textbf{ }$\left\vert j_{xy}\right\vert /$\textrm{P}$=1$\textbf{.}

\subsection{Minimizing the evolution time}

The starting point of the first approach can be summarized as follows. Given a
time-independent Hamiltonian \textrm{H} with a corresponding unitary
time-evolution operator $U\left(  t\right)  \overset{\text{def}}{=}%
e^{-\frac{i}{\hslash}\mathrm{H}t}$, one wishes to evolve a state $\left\vert
A\right\rangle $ into a state $\left\vert B\right\rangle $ in the shortest
possible time subject to the constraint that the difference $E_{+}-E_{-}$
between the largest ($E_{+}$) and the smallest ($E_{-}$) eigenvalues of
\textrm{H }is\textrm{ }kept fixed. In summary, we wish to find the optimal
Hamiltonian acting in the two-dimensional subspace spanned by $\left\vert
A\right\rangle $ and $\left\vert B\right\rangle $ that yields the optimal time
evolution subject to the constraint $E_{+}-E_{-}\overset{\text{def}}{=}%
E_{0}=\mathrm{fixed}$.

We begin by considering the following unitary evolution scheme,
\begin{equation}
\left\vert A\right\rangle =\binom{\alpha_{1}}{\alpha_{2}}\overset{e^{-\frac
{i}{\hslash}\mathrm{H}t}}{\rightarrow}\left\vert B\right\rangle =\binom
{\beta_{1}}{\beta_{2}}\text{, with }\mathrm{H}\overset{\text{def}}{=}\left(
\begin{array}
[c]{cc}%
h_{11} & h_{12}e^{-i\phi}\\
h_{12}e^{i\phi} & h_{22}%
\end{array}
\right)  \label{yo1}%
\end{equation}
and, where $h_{11}$, $h_{12}$, $h_{22}$, and $\phi$ are four real quantities.
We assume that $\left\vert A\right\rangle $ and $\left\vert B\right\rangle $
are normalized to one so that $\left\vert \alpha_{1}\right\vert ^{2}%
+\left\vert \alpha_{2}\right\vert ^{2}=1$ and $\left\vert \beta_{1}\right\vert
^{2}+\left\vert \beta_{2}\right\vert ^{2}=1$. The spectral decomposition of
$\mathrm{H}$ in Eq. (\ref{yo1})\ can be recast as $\mathrm{H}=E_{+}\left\vert
E_{+}\right\rangle \left\langle E_{+}\right\vert +E_{-}\left\vert
E_{-}\right\rangle \left\langle E_{-}\right\vert $ with the energy eigenvalue
constraint given by,%
\begin{equation}
\left(  E_{+}-E_{-}\right)  ^{2}=\left(  h_{11}-h_{22}\right)  ^{2}%
+4h_{12}^{2}=\mathrm{const}\overset{\text{def}}{=}E_{0}^{2}\text{.} \label{ec}%
\end{equation}
In view of our interest in studying the action of $e^{-\frac{i}{\hslash
}\mathrm{H}t}$ onto $\left\vert A\right\rangle $, it is convenient to observe
that $\mathrm{H}$ in Eq. (\ref{yo1}) can be decomposed in terms of the Pauli
matrices $\vec{\sigma}\overset{\text{def}}{=}\left(  \sigma_{x}\text{, }%
\sigma_{y}\text{, }\sigma_{z}\right)  $ as%
\begin{equation}
\mathrm{H}=\frac{h_{11}+h_{22}}{2}I+\frac{E_{0}}{2}\hat{a}\cdot\vec{\sigma
}\text{.} \label{yo2}%
\end{equation}
In Eq. (\ref{yo2}), $I$ denotes the identity matrix while $\hat{a}%
\overset{\text{def}}{=}$ $\vec{a}/\left\Vert \vec{a}\right\Vert $ with
$\left\Vert \vec{a}\right\Vert =E_{0}/2$ is a unit vector defined as,%
\begin{equation}
\hat{a}\overset{\text{def}}{=}\frac{2}{E_{0}}\left(  h_{12}\cos\left(
\phi\right)  \text{, }h_{12}\sin\left(  \phi\right)  \text{, }\frac
{h_{11}-h_{22}}{2}\right)  \text{.}%
\end{equation}
We emphasize at this stage that finding the optimal evolution Hamiltonian
reduces to finding the optimal set of the four real parameters $\left\{
h_{11},h_{12},h_{22},\phi\right\}  $ in\ Eq. (\ref{yo1}) or, equivalently, the
optimal $\hat{a}$ that appears in Eq. (\ref{yo2}). Using Eq. (\ref{yo2}) along
with algebraic manipulations that are typical in quantum mechanics with Pauli
matrices, the action of $e^{-\frac{i}{\hslash}\mathrm{H}t}$ onto $\left\vert
A\right\rangle $ leading to the state $\left\vert B\right\rangle $ in a time
$T_{AB}$ yields%
\begin{equation}
\binom{\beta_{1}}{\beta_{2}}=e^{-\frac{i}{\hslash}\frac{h_{11}+h_{22}}%
{2}T_{AB}}\binom{\alpha_{1}\left[  \cos\left(  \frac{E_{0}}{2\hslash}%
T_{AB}\right)  -i\frac{h_{11}-h_{22}}{E_{0}}\sin\left(  \frac{E_{0}}{2\hslash
}T_{AB}\right)  \right]  +\alpha_{2}\left[  -ie^{-i\phi}\frac{2h_{12}}{E_{0}%
}\sin\left(  \frac{E_{0}}{2\hslash}T_{AB}\right)  \right]  }{\alpha_{1}\left[
-ie^{i\phi}\frac{2h_{12}}{E_{0}}\sin\left(  \frac{E_{0}}{2\hslash}%
T_{AB}\right)  \right]  +\alpha_{2}\left[  \cos\left(  \frac{E_{0}}{2\hslash
}T_{AB}\right)  +i\frac{h_{11}-h_{22}}{E_{0}}\sin\left(  \frac{E_{0}}%
{2\hslash}T_{AB}\right)  \right]  }\text{.} \label{yo3}%
\end{equation}
At this point, we note that the components of the states $\left\vert
A\right\rangle $ and $\left\vert B\right\rangle $ depend on the choice of the
basis of the two-dimensional subspace spanned by these two vectors. Therefore,
for the sake of computational simplicity and without loss of generality, we
choose a basis so that $\left\vert A\right\rangle =\left(  1\text{, }0\right)
$ and $\left\vert B\right\rangle =\left(  \alpha\text{, }\beta\right)  $. With
this choice, Eq. (\ref{yo3}) reduces to%
\begin{equation}
\binom{\alpha}{\beta}=e^{-\frac{i}{\hslash}\frac{h_{11}+h_{22}}{2}T_{AB}%
}\binom{\cos\left(  \frac{E_{0}}{2\hslash}T_{AB}\right)  -i\frac{h_{11}%
-h_{22}}{E_{0}}\sin\left(  \frac{E_{0}}{2\hslash}T_{AB}\right)  }{-ie^{i\phi
}\frac{2h_{12}}{E_{0}}\sin\left(  \frac{E_{0}}{2\hslash}T_{AB}\right)
}\text{.} \label{yo4}%
\end{equation}
Considering the modulus of $\beta$ as expressed in Eq. (\ref{yo4}), we note
that the expression of the evolution time $T_{AB}$ becomes%
\begin{equation}
T_{AB}=\frac{2\hslash}{E_{0}}\sin^{-1}\left(  \frac{E_{0}\left\vert
\beta\right\vert }{2h_{12}}\right)  \text{.} \label{yo5}%
\end{equation}
We observe that the $\sin^{-1}\left(  x\right)  $ function is a monotonic
increasing function of its argument $x$ with $\sin^{-1}\left(  0\right)  =0$.
Therefore, since $E_{0}$ and $\left\vert \beta\right\vert $ are held fixed,
the minimum value of $T_{AB}$ in Eq. (\ref{yo5}) is reached when $h_{12}$
assumes its maximum possible value. The maximum possible value $h_{12}^{\max}$
of $h_{12}$ compatible with the eigenvalue constraint in Eq. (\ref{ec}) is
reached when $h_{11}=h_{22}$ and equals $h_{12}^{\max}=E_{0}/2$. Therefore,
the optimal evolution time $T_{AB}^{\min}=T_{AB}\left(  h_{12}^{\max}\right)
$ is equal to%
\begin{equation}
T_{AB}^{\min}=\frac{2\hslash}{E_{0}}\sin^{-1}\left(  \left\vert \beta
\right\vert \right)  \text{,} \label{yo6}%
\end{equation}
or, equivalently, $T_{AB}^{\min}=\left(  2\hslash/E_{0}\right)  \cos
^{-1}\left(  \left\vert \alpha\right\vert \right)  $ using the normalization
condition $\left\vert \alpha\right\vert ^{2}+\left\vert \beta\right\vert
^{2}=1$ along with properties of the $\sin^{-1}\left(  x\right)  $ function.
So far, we have realized that the optimal set of the four real parameters
$\left\{  h_{11},h_{12},h_{22},\phi\right\}  $ is specified by $h_{12}^{\max
}=E_{0}/2$ and $h_{11}=h_{22}$. Therefore, it remains to find the explicit
expressions for the optimal $h_{11}$ and $\phi$. These expressions can be
found as follows. Let us set $\alpha\overset{\text{def}}{=}\left\vert
\alpha\right\vert e^{i\varphi_{\alpha}}$ and $\beta\overset{\text{def}}%
{=}\left\vert \beta\right\vert e^{i\varphi_{\beta}}$ with $\varphi_{\alpha}$
and $\varphi_{\beta}$ in $%
\mathbb{R}
$. Inserting Eq. (\ref{yo6}) into Eq. (\ref{yo4}), we obtain
\begin{equation}
\binom{\left\vert \alpha\right\vert e^{i\varphi_{\alpha}}}{\left\vert
\beta\right\vert e^{i\varphi_{\beta}}}=e^{-\frac{i}{\hslash}h_{11}T_{AB}%
^{\min}}\binom{\sqrt{1-\left\vert \beta\right\vert ^{2}}}{-ie^{i\phi
}\left\vert \beta\right\vert }\text{.} \label{yo7}%
\end{equation}
With the help of some algebra, Eq. (\ref{yo7}) yields
\begin{equation}
h_{11}=-\frac{\omega_{0}}{2}\frac{\varphi_{\alpha}}{\sin^{-1}\left(
\left\vert \beta\right\vert \right)  }\text{, and }\phi=\varphi_{\beta
}-\varphi_{\alpha}+\frac{\pi}{2}\text{.} \label{yo8}%
\end{equation}
In conclusion, recalling the $h_{12}^{\max}=E_{0}/2$ and $h_{11}=h_{22}$
together with Eq. (\ref{yo8}), the optimal evolution Hamiltonian can be recast
as \cite{correction1}%
\begin{equation}
\mathrm{H}=\frac{E_{0}}{2}\left(
\begin{array}
[c]{cc}%
\frac{\varphi_{\alpha}}{\sin^{-1}\left(  \left\vert \beta\right\vert \right)
} & e^{-i\left(  \varphi_{\beta}-\varphi_{\alpha}-\frac{\pi}{2}\right)  }\\
e^{i\left(  \varphi_{\beta}-\varphi_{\alpha}-\frac{\pi}{2}\right)  } &
\frac{\varphi_{\alpha}}{\sin^{-1}\left(  \left\vert \beta\right\vert \right)
}%
\end{array}
\right)  \text{.} \label{hopt}%
\end{equation}
Interestingly, we note that the expectation value of the Hamiltonian
$\left\langle A|\mathrm{H|}A\right\rangle $ is equal to $\left(
\varphi_{\alpha}E_{0}\right)  /2\sin^{-1}\left(  \left\vert \beta\right\vert
\right)  $ while the energy uncertainty of $\mathrm{H}$ in Eq. (\ref{hopt}) is
given by $\Delta E\overset{\text{def}}{=}\left[  \left\langle A|\mathrm{H}%
^{2}\mathrm{|}A\right\rangle -\left\langle A|\mathrm{H|}A\right\rangle
^{2}\right]  ^{1/2}=E_{0}/2$. Using the energy eigenvalue constraint in Eq.
(\ref{ec}), we also notice that $\Delta E=\left(  E_{+}-E_{-}\right)  /2$. As
a final remark, we point out that since overall phases of state vectors have
unobservable effects in quantum mechanics, the Hamiltonians \textrm{H} and
\textrm{H}$-(1/2)\mathrm{tr}\left(  \mathrm{H}\right)  I$ assume an identical
maximal value $\Delta E_{\max}$ of energy uncertainty $\Delta E$. Therefore,
despite having different expectation values, these Hamiltonians generate the
same physics of quantum evolutions. For this reason, for example, one may set
the phase $\varphi_{\alpha}$ in Eq.(\ref{hopt}) equal to zero.

Starting from a traceless Hamiltonian with $\Delta E_{\max}=\left(
E_{+}-E_{-}\right)  /2$ will be the starting point in the second approach to
optimal quantum evolutions that we treat in our paper. This second approach is
based upon maximizing the energy uncertainty rather than minimizing the time
evolution and will be discussed in the next subsection.

\subsection{Maximizing the energy uncertainty}

The starting point of the second approach can be summarized as follows.
Consider a time-independent and traceless Hamiltonian \textrm{H }with a
spectral decomposition given by \textrm{H}$=E_{-}\left\vert E_{-}\right\rangle
\left\langle E_{-}\right\vert +E_{+}\left\vert E_{+}\right\rangle \left\langle
E_{+}\right\vert $, where $E_{+}\geq E_{-}$ and $\left\langle E_{+}%
|E_{-}\right\rangle =\delta_{+,-}$. One wishes to evolve a state (not
necessarily normalized) $\left\vert A\right\rangle $ into a state $\left\vert
B\right\rangle $ in the shortest possible time by maximizing the energy
uncertainty $\Delta E\overset{\text{def}}{=}\left[  \left\langle
A|\mathrm{H}^{2}\mathrm{|}A\right\rangle /\left\langle A|A\right\rangle
-\left(  \left\langle A|\mathrm{H|}A\right\rangle /\left\langle
A|A\right\rangle \right)  ^{2}\right]  ^{1/2}$ and obtain $\Delta E=\Delta
E_{\max}$.

For the sake of simplicity, we denote $\left\vert E_{\pm}\right\rangle
=\left\vert E_{2,1}\right\rangle $ and $E_{\pm}=E_{2,1}$ in what follows. To
find the value of $\Delta E_{\max}$, we note that an arbitrary unnormalized
initial state $\left\vert A\right\rangle $ can be decomposed as $\left\vert
A\right\rangle =\alpha_{1}\left\vert E_{1}\right\rangle +\alpha_{2}\left\vert
E_{2}\right\rangle $ with $\alpha_{1}$, $\alpha_{2}\in%
\mathbb{C}
$. Then, after some algebra, we get%
\begin{equation}
\Delta E=\frac{E_{2}-E_{1}}{2}\left[  1-\left(  \frac{\left\vert \alpha
_{1}\right\vert ^{2}-\left\vert \alpha_{2}\right\vert ^{2}}{\left\vert
\alpha_{1}\right\vert ^{2}+\left\vert \alpha_{2}\right\vert ^{2}}\right)
^{2}\right]  ^{1/2}\text{.} \label{chi4}%
\end{equation}
From Eq. (\ref{chi4}), we note that the maximum value of $\Delta E$ is
obtained for $\left\vert \alpha_{1}\right\vert =\left\vert \alpha
_{2}\right\vert $ (with $\alpha_{1}$ and $\alpha_{2}$ given by $\left\langle
E_{1}|A\right\rangle $ and $\left\langle E_{2}|A\right\rangle $, respectively)
and equals $\Delta E_{\max}\overset{\text{def}}{=}\left(  E_{2}-E_{1}\right)
/2$ as mentioned in the previous subsection. The main underlying idea in this
second approach is that of recasting \textrm{H}$=E_{1}\left\vert
E_{1}\right\rangle \left\langle E_{1}\right\vert +E_{2}\left\vert
E_{2}\right\rangle \left\langle E_{2}\right\vert $ in terms of the initial and
final states $\left\vert A\right\rangle $ and $\left\vert B\right\rangle $
while keeping,
\begin{equation}
\Delta E=\Delta E_{\max}\overset{\text{def}}{=}\frac{E_{2}-E_{1}}{2}\text{.}
\label{ec2}%
\end{equation}
Observe that in terms of the eigenvectors of the Hamiltonian, $\left\vert
A\right\rangle $ and $\left\vert B\right\rangle $ can be decomposed as
$\left\vert A\right\rangle =\alpha_{1}\left\vert E_{1}\right\rangle
+\alpha_{2}\left\vert E_{2}\right\rangle $, and $\left\vert B\right\rangle
=\beta_{1}\left\vert E_{1}\right\rangle +\beta_{2}\left\vert E_{2}%
\right\rangle $, respectively. To ensure minimum travel time $T_{AB}^{\min}$
(that is, $\Delta E=\Delta E_{\max}$), we need to set $\left\vert \alpha
_{1}\right\vert =\left\vert \alpha_{2}\right\vert $ and $\left\vert \beta
_{1}\right\vert =\left\vert \beta_{2}\right\vert $. Therefore, let $\alpha
_{2}=e^{i\varphi_{\alpha}}\alpha_{1}$ and $\beta_{2}=e^{i\varphi_{\beta}}%
\beta_{1}$ with $\varphi_{\alpha,\beta}\in%
\mathbb{R}
$. Then, $\left\vert A\right\rangle $ and $\left\vert B\right\rangle $ become%
\begin{equation}
\left\vert A\right\rangle =\alpha_{1}\left\vert E_{1}\right\rangle +\alpha
_{2}\left\vert E_{2}\right\rangle =\alpha_{1}\left\vert E_{1}\right\rangle
+e^{i\varphi_{\alpha}}\alpha_{1}\left\vert E_{2}\right\rangle \text{,}
\label{chi5a}%
\end{equation}
and,%
\begin{equation}
\left\vert B\right\rangle =\beta_{1}\left\vert E_{1}\right\rangle +\beta
_{2}\left\vert E_{2}\right\rangle =\beta_{1}\left\vert E_{1}\right\rangle
+e^{i\varphi_{\beta}}\beta_{1}\left\vert E_{2}\right\rangle \text{,}
\label{chi5b}%
\end{equation}
respectively. From Eqs. (\ref{chi5a}) and (\ref{chi5b}), we obtain $\left\vert
E_{1}\right\rangle +e^{i\varphi_{\alpha}}\left\vert E_{2}\right\rangle
=\alpha_{1}^{-1}\left\vert A\right\rangle \overset{\text{def}}{=}\sqrt
{2}\left\vert \mathcal{A}\right\rangle $ and $\left\vert E_{1}\right\rangle
+e^{i\varphi_{\beta}}\left\vert E_{2}\right\rangle =\beta_{1}^{-1}\left\vert
B\right\rangle \overset{\text{def}}{=}\sqrt{2}e^{-i\frac{\varphi_{\alpha
}-\varphi_{\beta}}{2}}\left\vert \mathcal{B}\right\rangle $. After some matrix
algebra, we get%
\begin{equation}
\left(
\begin{array}
[c]{c}%
\left\vert E_{1}\right\rangle \\
\left\vert E_{2}\right\rangle
\end{array}
\right)  =\frac{\sqrt{2}}{e^{i\frac{\varphi_{\alpha}+\varphi_{\beta}}{2}%
}-e^{i\varphi_{\alpha}}e^{i\frac{\varphi_{\alpha}-\varphi_{\beta}}{2}}}\left(
\begin{array}
[c]{cc}%
e^{i\frac{\varphi_{\alpha}+\varphi_{\beta}}{2}} & -e^{i\varphi_{\alpha}}\\
-e^{i\frac{\varphi_{\alpha}-\varphi_{\beta}}{2}} & 1
\end{array}
\right)  \left(
\begin{array}
[c]{c}%
\frac{\alpha_{1}^{-1}}{\sqrt{2}}\left\vert A\right\rangle \\
\frac{\beta_{1}^{-1}}{\sqrt{2}}e^{i\frac{\varphi_{\alpha}-\varphi_{\beta}}{2}%
}\left\vert B\right\rangle
\end{array}
\right)  \text{.} \label{chichi}%
\end{equation}
For the sake of completeness, we emphasize that%
\begin{equation}
\left\vert \left\langle \mathcal{A}|\mathcal{B}\right\rangle \right\vert
^{2}=\frac{\left\vert \left\langle A|B\right\rangle \right\vert ^{2}%
}{\left\langle A|A\right\rangle \left\langle B|B\right\rangle }=\cos
^{2}\left(  \frac{\varphi_{\alpha}-\varphi_{\beta}}{2}\right)  =\cos
^{2}\left(  \frac{\theta}{2}\right)  \text{,}%
\end{equation}
with $\theta\overset{\text{def}}{=}\varphi_{\alpha}-\varphi_{\beta
}=2s_{\text{\textrm{FS}}}=s_{\text{\textrm{geo}}}$ where $s_{\text{\textrm{FS}%
}}$ and $s_{\text{\textrm{geo}}}$ denote the Fubini-Study and the geodesic
distances, respectively. Finally, using Eq. (\ref{chichi}) along with noting
that $E_{2}=-E_{1}\overset{\text{def}}{=}E$ since the Hamiltonian is assumed
to be traceless, we obtain after some simple but tedious algebra that the
spectral decomposition \textrm{H}$=E_{1}\left\vert E_{1}\right\rangle
\left\langle E_{1}\right\vert +E_{2}\left\vert E_{2}\right\rangle \left\langle
E_{2}\right\vert $ becomes \cite{correction2}%
\begin{equation}
\mathrm{H}=\frac{iE}{\sin\left(  \frac{\varphi_{\alpha}-\varphi_{\beta}}%
{2}\right)  }\left[  \left\vert \mathcal{B}\right\rangle \left\langle
\mathcal{A}\right\vert -\left\vert \mathcal{A}\right\rangle \left\langle
\mathcal{B}\right\vert \right]  \text{.} \label{Hami}%
\end{equation}
In terms of the original initial and final states $\left\vert A\right\rangle $
and $\left\vert B\right\rangle $, after some additional simple but laborious
algebra, the Hamiltonian in Eq. (\ref{Hami}) can be finally recast as%
\begin{equation}
\mathrm{H}=iE\cot\left(  \frac{\varphi_{\alpha}-\varphi_{\beta}}{2}\right)
\left[  \frac{\left\vert B\right\rangle \left\langle A\right\vert
}{\left\langle A|B\right\rangle }-\frac{\left\vert A\right\rangle \left\langle
B\right\vert }{\left\langle B|A\right\rangle }\right]  \text{,} \label{amy}%
\end{equation}
while the geodesic line $\left\vert \psi\left(  t\right)  \right\rangle
=e^{-\frac{i}{\hslash}\mathrm{H}t}\left\vert A\right\rangle $ with
$\mathrm{H}$ in Eq. (\ref{amy})\ connecting the two states $\left\vert
A\right\rangle $ and $\left\vert B\right\rangle $ can be written as%
\begin{equation}
\left\vert \psi\left(  t\right)  \right\rangle =\left[  \cos\left(  \frac
{E}{\hslash}t\right)  -\frac{\cos\left(  \frac{\varphi_{\alpha}-\varphi
_{\beta}}{2}\right)  }{\sin\left(  \frac{\varphi_{\alpha}-\varphi_{\beta}}%
{2}\right)  }\sin\left(  \frac{E}{\hslash}t\right)  \right]  \left\vert
A\right\rangle +\frac{e^{i\frac{\varphi_{\alpha}-\varphi_{\beta}}{2}}}%
{\sin\left(  \frac{\varphi_{\alpha}-\varphi_{\beta}}{2}\right)  }\sin\left(
\frac{E}{\hslash}t\right)  \left\vert B\right\rangle \text{, }
\label{geodesic}%
\end{equation}
where $0\leq t\leq T_{AB}^{\min}$ with $T_{AB}^{\min}=\hslash\theta/\left(
2E\right)  $. For the sake of completeness, we note that for $\mathrm{H}$ in
Eq. (\ref{amy}), we correctly get $\left\langle A|\mathrm{H|}A\right\rangle
/\left\langle A|A\right\rangle =0$ and $\Delta E=\left[  \left\langle
A|\mathrm{H}^{2}\mathrm{|}A\right\rangle /\left\langle A|A\right\rangle
\right]  ^{1/2}=E=\Delta E_{\max}$. In conclusion, the Hamiltonians in Eqs.
(\ref{hopt}) and (\ref{amy}) are optimal-speed Hamiltonians yielding unit
quantum geometric efficiency $\eta_{\mathrm{QM}}=1$. For clarity, we emphasize
that Hamiltonians in Eqs. (\ref{hopt}) and (\ref{amy}) are both optimal speed
Hamiltonians. However, the Hamiltonian $\mathrm{H}$\textbf{ }in Eq.
(\ref{hopt}) is not traceless and its applicability is formally limited to
connecting the initial state $\left\vert A\right\rangle =\left(  1\text{,
}0\right)  $ to an arbitrary final state $\left\vert B\right\rangle $. The
Hamiltonian $\mathrm{H}$ in Eq. (\ref{amy}), instead, is traceless and can
connect an arbitrary initial state $\left\vert A\right\rangle $ to an
arbitrary final state $\left\vert B\right\rangle $.

Having discussed in detail the two main constructions of Hamiltonians yielding
unit quantum geometric efficiency, in the next section we focus on the
geometric characterization of the propagation of polarized light with maximal
degree of coherence. As we present this classical optics description, we will
emphasize analogies and determine exactly correspondences with the above
mentioned quantum mechanical characterizations.

\section{Propagation of polarized light and degree of coherence}

In this section, we describe the propagation of light by means of the
polarization ellipse, the Stokes parameters, and the Poincar\'{e} sphere.
Then, we briefly define the notions of coherence of electric vibrations,
degree of coherence, coherency matrix, and degree of polarization of a light
wave. We end this section with a discussion on propagation of polarized light
with maximal degree of coherence, that is, unit classical optical efficiency
on the Poincar\'{e} sphere. However, before beginning with our formal
descriptions, we present some motivational background that helps explaining
our underlying motivations for presenting this material of polarization optics.

\subsection{Motivational background}

Two important quantities in optics when studying the physics of polarized
light are the degree of polarization \textrm{P} of the wave and the degree of
coherence $\left\vert j_{xy}\right\vert $ of the electric vibrations. The
quantity \textrm{P }is defined as \cite{born50},%
\begin{equation}
\mathrm{P}\overset{\text{def}}{=}\frac{I_{\mathrm{pol}}}{I_{\mathrm{tot}}%
}\text{,} \label{polarization}%
\end{equation}
with $I_{\mathrm{tot}}\overset{\text{def}}{=}$ $I_{\mathrm{pol}}%
+I_{\mathrm{unpol}}$ denoting the total intensity of the wave and
$I_{\mathrm{pol}}$ being the intensity of the monochromatic (hence, polarized)
part of the wave. The quantity $\mathrm{P}$ with $0\leq\mathrm{P}\leq1$
expresses the \textquotedblleft amount of polarization\textquotedblright%
\ present in the wave. In particular, the wave is completely unpolarized when
$\mathrm{P}=0$ and completely polarized when $\mathrm{P}=1$. When
$0<\mathrm{P}<1$, the light is partially polarized. The degree of coherence
$\left\vert j_{xy}\right\vert $, instead, is defined as the modulus of the
complex degree of coherence $j_{xy}$ \cite{born50},%
\begin{equation}
j_{xy}\overset{\text{def}}{=}\frac{J_{xy}}{\sqrt{J_{xx}}\sqrt{J_{yy}}}\text{.}
\label{jec}%
\end{equation}
In Eq. (\ref{jec}), $J_{ij}\overset{\text{def}}{=}\left\langle E_{i}\left(
t\right)  E_{j}^{\ast}\left(  t\right)  \right\rangle $ are the matrix
coefficients of the so-called coherency matrix $J$ and the angle brackets
denote the time average operation. The quantity $\left\vert j_{xy}\right\vert
$ with $0\leq\left\vert j_{xy}\right\vert \leq\left\vert j_{xy}\right\vert
_{\max}$, instead, measures the degree of correlation of the electric
vibrations. When $\left\vert j_{xy}\right\vert =0$, the electric vibrations
are uncorrelated and they may be said to be incoherent. Furthermore, when
$\left\vert j_{xy}\right\vert =\left\vert j_{xy}\right\vert _{\max}$, the
vibrations may be said to be coherent. Finally, when $0<\left\vert
j_{xy}\right\vert <\left\vert j_{xy}\right\vert _{\max}$, vibrations are known
as partially coherent\textbf{.} The quantity $\mathrm{P}$ can be fully
expressed in terms of the determinant and the trace of the coherency matrix
$J$ as shown in Ref. \cite{born50}. Therefore, it is a quantity whose value
does not change under arbitrary rotations of the orthogonal Cartesian axes
used to describe the electric vibrations. However, unlike the degree of
polarization of the wave, the degree of coherence $\left\vert j_{xy}%
\right\vert $ between the electric vibrations in any two mutually orthogonal
directions of propagation of the wave is generally affected by the specific
choice of the two orthogonal directions \cite{born50}. In particular, it is
possible to show that there always exist a pair of orthogonal directions for
which the degree of coherence $\left\vert j_{xy}\right\vert $ of the electric
vibrations reaches its maximum value $\left\vert j_{xy}\right\vert _{\max}$
and, in addition, this value equals the degree of polarization $\mathrm{P}$ of
the wave \cite{wolf59}. Interestingly, this particular pair of orthogonal
directions has a clear geometrical interpretation. Indeed, representing the
wave as an incoherent mixture of a wave of natural radiation and a wave of
monochromatic (hence, completely polarized) radiation, it can be shown that
these directions for which the degree of coherence $\left\vert j_{xy}%
\right\vert $ equals the degree of polarization $\mathrm{P}$ are the bisectors
of the principal directions (that is, major and minor axes) of the
polarization ellipse of the polarized portion of the wave \cite{chandra50}.

The existence of a pair of directions $\left(  \hat{x}^{\prime}\text{, }%
\hat{y}^{\prime}\right)  $ rotated around the $\hat{z}$-axis (that is, the
axis that specifies the direction of propagation of the wave) by a specific
angle $\varphi_{\mathrm{opt}}$ that affects the electric vibrations in such a
manner that the quantity $\eta_{\mathrm{opt}}\overset{\text{def}}{=}\left\vert
j_{xy}\right\vert /\mathrm{P}$, that we name \emph{classical optical
efficiency} in this paper, equals one is reminiscent of the existence of a
pair of orthogonal states $\left(  \left\vert E_{+}\right\rangle \text{,
}\left\vert E_{-}\right\rangle \right)  $ that defines the axis of rotation of
the Bloch sphere orthogonal to the hemispherical plane containing the initial
and final unit states $\left\vert A\right\rangle $ and $\left\vert
B\right\rangle $ as discussed in the previous section. In the quantum case,
this rotation around the $\hat{n}_{E_{+}}$-axis by an angle $2\cos^{-1}\left[
\left\vert \left\langle A|B\right\rangle \right\vert \right]  $ is essentially
the unitary evolution operator emerging from the optimal-speed Hamiltonian
that yields efficiency $\eta_{\mathrm{QM}}\overset{\text{def}}{=}s_{0}/s$
equal to one. Furthermore, just as these directions $\left(  \hat{x}^{\prime
}\text{, }\hat{y}^{\prime}\right)  $ for which $\left\vert j_{xy}\right\vert
=\mathrm{P}$ are the bisectors of the principal directions $\left(  \hat{\xi
}\text{, }\hat{\eta}\right)  $ corresponding to the major and minor axes,
respectively, of the polarization ellipse, in a similar fashion, the pair of
orthogonal states $\left(  \left\vert E_{+}\right\rangle \text{, }\left\vert
E_{-}\right\rangle \right)  $ for which $s_{0}=s$ lie in the equatorial plane
when the pair of states $\left(  \left\vert A\right\rangle \text{, }\left\vert
A_{\bot}\right\rangle \right)  $ are assumed to lie at the poles. Finally, the
angle $\varphi_{\mathrm{opt}}$ seems to be replaced in the quantum case by the
azimuthal angle $\varphi_{E_{+}}$ that serves to specify the location of
$\left\vert E_{+}\right\rangle $ on the Bloch sphere. We shall devote the rest
of this paper to make these formal analogies as quantitative as possible. We
shall begin by observing that to better characterize the propagation of the
effects of this simple two-dimensional rotation of the canonical Cartesian
axes $\left(  \hat{x}\text{, }\hat{y}\right)  $ on the electric vibrations
along with the coherency matrix $J$ and, ultimately, on the degree of
coherence $\left\vert j_{xy}\right\vert $, we need to better understand how to
visualize and perform calculations when considering polarized light. For this
reason, we shall introduce the concepts of polarization ellipse along with
that of the Poincar\'{e} sphere.

To better motivate the definitions and concepts of polarization optics in what
follows, we present in Table I a schematic depiction of the most relevant
quantities that specify optimal-speed unitary quantum time evolutions on the
Bloch sphere together with the analog quantities that characterize the
propagation of light with maximal degree of coherence by means of the
polarization ellipse and the Poincar\'{e} e sphere representations of
polarized light. Clearly, these correspondences will become more transparent
as we go through the next subsections and Sec. IV.\begin{table}[t]
\centering
\begin{tabular}
[c]{c|c|c}\hline\hline
\textbf{Bloch sphere} & \textbf{Polarization ellipse} & \textbf{Poincar\'{e}
sphere}\\\hline
$\left(  \left\vert A\right\rangle \text{, }\left\vert A_{\perp}\right\rangle
\right)  $ & $\left(  \hat{\xi}\text{, }\hat{\eta}\right)  $ & $\vec
{S}_{\text{\textrm{initial}}}$\\
$\left(  \left\vert E_{+}\right\rangle _{\mathrm{sub-optimal}}\text{,
}\left\vert E_{-}\right\rangle _{\mathrm{sub-optimal}}\right)  $ & $\left(
\hat{x}\text{, }\hat{y}\right)  $ & $\vec{S}_{\mathrm{sub-optimal}}$\\
$\left(  \left\vert E_{+}\right\rangle _{\mathrm{optimal}}\text{, }\left\vert
E_{-}\right\rangle _{\mathrm{optimal}}\right)  $ & $\left(  \hat{x}^{\prime
}\text{, }\hat{y}^{\prime}\right)  $ & $\vec{S}_{\mathrm{optimal}}$\\
$\mathcal{H}=a_{0}I+\vec{a}\cdot\vec{\sigma}$ & $\vec{E}$ & $J=\frac{1}{2}%
\vec{S}\cdot\vec{\sigma}$\\
$\hat{a}=\hat{a}\left(  \theta\text{, }\varphi\right)  $ & $\vec{E}=\vec
{E}\left(  \beta\text{, }\chi\right)  $ & $\vec{S}=\vec{S}\left(
2\beta\text{, }2\chi\right)  $\\
$e^{-i\frac{\left\Vert \vec{a}\right\Vert T_{AB}}{\hslash}\hat{a}\cdot
\vec{\sigma}}$ & \textrm{R}$_{\hat{z}}\left(  \alpha\right)  $ &
\textrm{M}$_{\mathrm{ROT}}\left(  \alpha\right)  $\\
$s_{0}$, fixed & $\vec{E}$, fixed & $\mathrm{P}$, fixed\\
$s\left(  0\right)  \rightarrow s\left(  T_{AB}\right)  $ & $\left[  \vec
{E}\right]  _{\left\{  \hat{x}\text{, }\hat{y}\right\}  }\rightarrow\left[
\vec{E}\right]  _{\left\{  \hat{x}^{\prime}\text{, }\hat{y}^{\prime}\right\}
}$ & $\left\vert j_{xy}\right\vert \rightarrow\left\vert j_{x^{\prime
}y^{\prime}}\right\vert $\\
$\eta_{\mathrm{QM}}\overset{\text{def}}{=}s_{0}/s$ & $\left\langle E_{x}%
E_{x}^{\ast}\right\rangle -\left\langle E_{y}E_{y}^{\ast}\right\rangle $ &
$\eta_{\mathrm{optics}}\overset{\text{def}}{=}\left\vert j_{xy}\right\vert
/\mathrm{P}$\\\hline
\end{tabular}
\caption{Schematic depiction of the most relevant quantities that specify
optimal-speed unitary quantum time evolutions on the Bloch sphere together
with the analog quantities that characterize the propagation of light with
maximal degree of coherence by means of the polarization ellipse and the
Poincar\'{e} sphere representations of polarized light.}%
\end{table}

\subsection{Polarization of a light wave}

In the previous section, we have explained how the Hamiltonian operator
affects the path of evolution of a quantum system (specifically, a spin-$1/2$
particle) in terms of geometric evolutions on a Bloch sphere. In this section,
keeping the directions of rays of light constant during its propagation, we
focus on the state of polarization and the intensity of the light as it passes
through an optical system. In this case, the three fundamental types of
optical elements are wave plates, rotators, and polarizers. These elements
give rise to phase shifting, rotations, and anisotropic attenuation,
respectively. More specifically, we are interested here in
intensity-preserving linear optical transformations which quantify the effect
of rotators on polarized light in a geometric fashion. For such a
quantification, we need to arrive at the Poincar\'{e} sphere description of
polarized light. The way we plan to pursue this goal can be outlined as
follows. First, we begin with the polarization ellipse representation of
polarized light \cite{born50}. Second, we introduce the Stokes parameters from
the polarization ellipse \cite{collett68}. Finally, we introduce the
Poincar\'{e} sphere by attaching a geometric interpretation to the Stokes
parameters \cite{walker54}. We remark that the traditional language for
studying the two-component electric vector of the light is the so-called
Jones-matrix formalism based upon the use of $2\times2$ complex matrices
\cite{jones41}. Alternatively, regarding the Stokes parameters as the
components of a column matrix or four-vector and optical devices as
represented by $4\times4$ matrices \cite{mueller48}, the so-called Mueller
matrix method can be employed to quantify the effect of optical devices on
polarized light. For further details on the Mueller matrices in optics, we
refer to Appendix A.

\emph{Polarization ellipse}. Assume that the electric vector field $\vec{E}$
of the light propagating along the $\hat{z}$-axis is given by $\vec{E}%
=E_{x}\left(  t\right)  \hat{x}+E_{y}\left(  t\right)  \hat{y}$ with
$E_{x}\left(  t\right)  $ and $E_{y}\left(  t\right)  $ defined as
$E_{x}\left(  t\right)  \overset{\text{def}}{=}E_{0x}\left(  t\right)
\cos\left[  \omega t+\delta_{x}\left(  t\right)  \right]  $ and $E_{y}\left(
t\right)  \overset{\text{def}}{=}E_{0y}\left(  t\right)  \cos\left[  \omega
t+\delta_{y}\left(  t\right)  \right]  $, respectively. The quantities
$\omega$, $\delta_{x}\left(  t\right)  $, and $\delta_{y}\left(  t\right)  $
specify the plane wave and denote the instantaneous angular frequency and the
two instantaneous phases, respectively. After some algebraic manipulations of
the two relations involving $E_{x}\left(  t\right)  $ and $E_{y}\left(
t\right)  $, one arrives at an equation of an ellipse in a nonstandard form
given by
\begin{equation}
\frac{E_{x}^{2}\left(  t\right)  }{E_{0x}^{2}\left(  t\right)  }+\frac
{E_{y}^{2}\left(  t\right)  }{E_{0y}^{2}\left(  t\right)  }-\frac
{2E_{x}\left(  t\right)  E_{y}\left(  t\right)  }{E_{0x}\left(  t\right)
E_{0y}\left(  t\right)  }\cos\left[  \delta\left(  t\right)  \right]
=\sin^{2}\left[  \delta\left(  t\right)  \right]  \text{,} \label{ellipse}%
\end{equation}
with $\delta\overset{\text{def}}{=}$ $\delta_{x}-\delta_{y}$ \cite{collett68}.
The ellipse defined by Eq. (\ref{ellipse}) is not in its standard from since
$E_{x}\left(  t\right)  $ and $E_{y}\left(  t\right)  $ are not directed along
the $\hat{x}$- \ and $\hat{y}$-axes. Instead, they are directed along the
$\hat{\xi}$- and $\hat{\eta}$-directions obtained from the canonical Cartesian
axes via a rotation around the $\hat{z}$-axis by an angle $\chi$. This angle
is known as the orientation angle with $0\leq\chi<\pi$ and, clearly, it
describes how tilted is the ellipse with respect to the canonical Cartesian
axes. For the sake of completeness and later use, we also introduce at this
point the so-called ellipticity angle $\beta$ with $-\pi/4<\beta\leq\pi/4$
defined as $\tan\beta\overset{\text{def}}{=}b/a$ with $a$ and $b$ being the
major and minor axes of the polarization ellipse, respectively. This angle
specifies the shape of the ellipse. In what follows, we introduce the Stokes
parameters from the polarization ellipse.

\emph{The Stokes parameters from the polarization ellipse}. Focusing on
monochromatic radiation with $E_{0x}$, $E_{0y}$, $\delta_{x}$, and $\delta
_{y}$ constant in time, Eq. (\ref{ellipse}) reduces to%
\begin{equation}
\frac{E_{x}^{2}\left(  t\right)  }{E_{0x}^{2}}+\frac{E_{y}^{2}\left(
t\right)  }{E_{0y}^{2}}-\frac{2E_{x}\left(  t\right)  E_{y}\left(  t\right)
}{E_{0x}E_{0y}}\cos\delta=\sin^{2}\delta\text{.} \label{ellipse1}%
\end{equation}
To represent Eq. (\ref{ellipse1}) in terms of observables of the
electromagnetic radiation, one needs to consider a time average over an
infinite time interval. However, given the periodic behavior of $E_{x}\left(
t\right)  $ and $E_{y}\left(  t\right)  $, averaging over a single period of
vibration $T$ will suffice. Specifically, define the time average of
$E_{i}\left(  t\right)  E_{j}\left(  t\right)  $ as%
\begin{equation}
\left\langle E_{i}\left(  t\right)  E_{j}\left(  t\right)  \right\rangle
\overset{\text{def}}{=}\frac{1}{T}\int_{0}^{T}E_{i}\left(  t\right)
E_{j}\left(  t\right)  dt\text{.} \label{timeaverage}%
\end{equation}
Using Eq. (\ref{timeaverage}), it can be shown following Ref. \cite{collett68}
that the time-averaged version of Eq. (\ref{ellipse1}) can be recast as%
\begin{equation}
S_{0}^{2}=S_{1}^{2}+S_{2}^{2}+S_{3}^{2}\text{,} \label{completely1}%
\end{equation}
with $S_{0}\overset{\text{def}}{=}E_{0x}^{2}+E_{0y}^{2}$, $S_{1}%
\overset{\text{def}}{=}E_{0x}^{2}-E_{0y}^{2}$, $S_{2}\overset{\text{def}}%
{=}2E_{0x}E_{0y}\cos\delta$, and $S_{3}\overset{\text{def}}{=}2E_{0x}%
E_{0y}\sin\delta$. The four parameters $\left\{  S_{i}\right\}  $ with $0\leq
i\leq3$ are the observables of the polarization ellipse with $S_{0}$ being the
total intensity of the radiation while $\left\{  S_{1}\text{, }S_{2}\text{,
}S_{3}\right\}  $ specify the state of polarization of the light beam. These
are the so-called four Stokes polarization parameters \cite{stokes52}.
Equation (\ref{completely1}) holds for completely polarized light and $S_{0}$
is redundant in this case. Instead, for partially polarized light, $S_{0}%
^{2}\geq S_{1}^{2}+S_{2}^{2}+S_{3}^{2}$ and $S_{0}$ is no longer redundant.
The excess $S_{0}^{2}-\left(  S_{1}^{2}+S_{2}^{2}+S_{3}^{2}\right)  $
indicates the amount of unpolarized light present in the beam. More\textbf{
}specifically, for completely polarized light beams\textbf{, }$S_{0}%
\overset{\text{def}}{=}I_{\mathrm{tot}}=I_{\mathrm{pol}}$\textbf{.} Instead,
for partially polarized light\textbf{, }$S_{0}\overset{\text{def}}%
{=}I_{\mathrm{tot}}>I_{\mathrm{pol}}$\textbf{. }We refer to Appendix B for
details on the behavior of $\left\vert j_{xy}\right\vert $ for partially
polarized waves using the Poincar\'{e} sphere formalism.\textbf{ }

In what follows, we introduce the Poincar\'{e} sphere by attaching a geometric
interpretation to the Stokes parameters.

\emph{The Poincar\'{e} sphere from the Stokes parameters}. It can be verified
by a straightforward but tedious computation as mentioned in Refs.
\cite{walker54,perrin42} that for a fixed value of $S_{0}$, we have for
completely polarized light $S_{1}\overset{\text{def}}{=}S_{0}\cos\left(
2\beta\right)  \cos\left(  2\chi\right)  $, $S_{2}\overset{\text{def}}{=}%
S_{0}\cos\left(  2\beta\right)  \sin\left(  2\chi\right)  $, and
$S_{3}\overset{\text{def}}{=}S_{0}\sin\left(  2\beta\right)  $ with $\beta$
and $\chi$ being the ellipticity and orientation angles, respectively, as
previously defined. Setting $S_{0}=1$, the quantities $\left\{  S_{1}\text{,
}S_{2}\text{, }S_{3}\right\}  $ have the following geometric interpretation.
Consider the vector $\vec{s}\overset{\text{def}}{=}\left(  S_{1}\text{, }%
S_{2}\text{, }S_{3}\right)  $ with length $\left\Vert \vec{s}\right\Vert
=S_{0}=1$. The vector $\vec{s}$ is located on a sphere of unit length with its
location determined by the azimuth angle $2\chi$ and the latitude angle
$2\beta$. Thus, a beam of elliptically polarized light can be specified by the
vector $\vec{s}$ an mapped on the sphere as originally pointed out by
Poincar\'{e} in Ref. \cite{poincare92}. For a graphical depiction of the Bloch
and Poincar\'{e} spheres, we refer to Fig. $1$. For more details on the
parametrization of qubits and polarization states viewed as points on the
Bloch sphere and the Poincar\'{e} sphere, respectively, we refer to Appendix
C. \begin{figure}[t]
\centering
\includegraphics[width=0.60\textwidth] {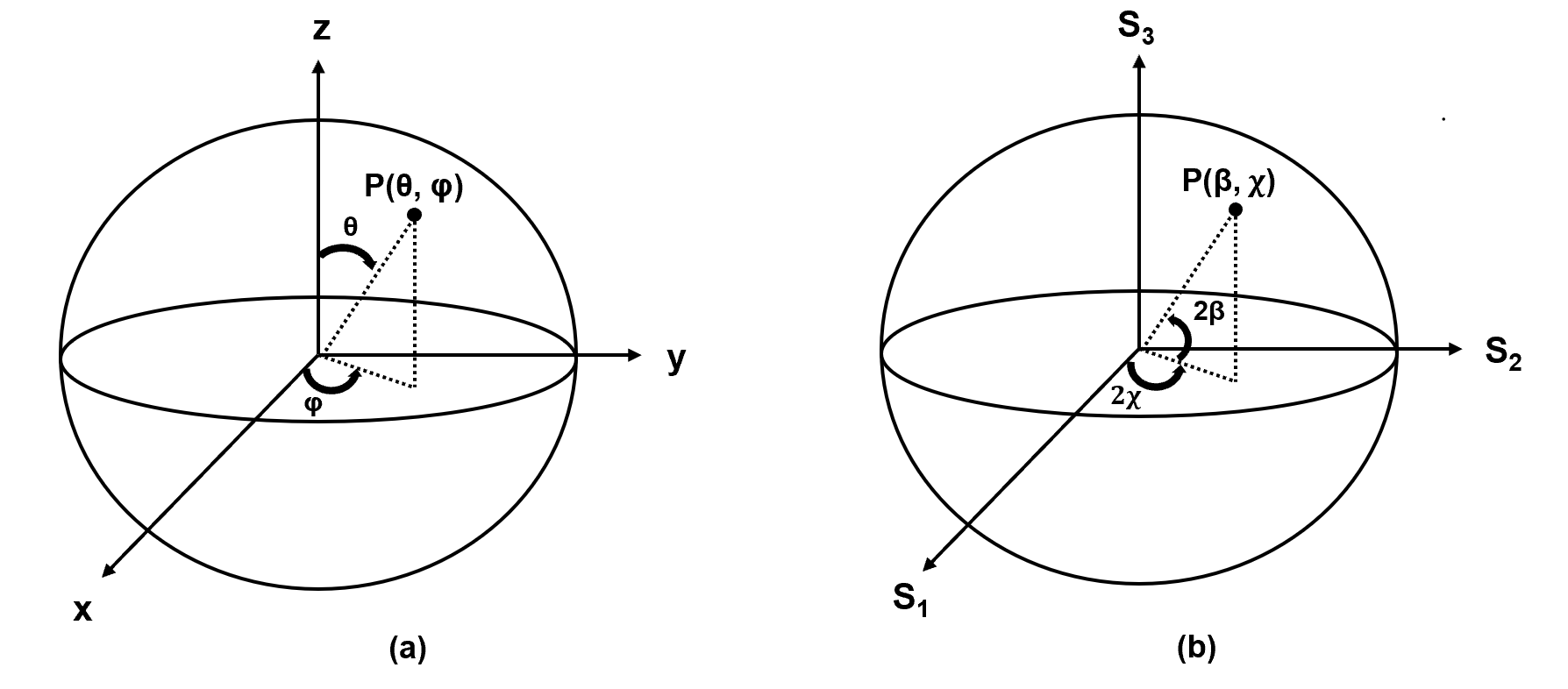}\caption{(a) The Bloch sphere for
pure quantum states of a single qubit. A point $P=P\left(  \theta\text{,
}\varphi\right)  $ on the surface of the Bloch sphere is defined by the Bloch
vector $\vec{r}\overset{\text{def}}{=}\left(  r_{x}\text{, }r_{y}\text{,
}r_{z}\right)  =\left(  r\sin\theta\cos\varphi\text{, }r\sin\theta\sin
\varphi\text{, }r\cos\theta\right)  $ with $\left\Vert \vec{r}\right\Vert =1$.
Mixed states are specified by $\left\Vert \vec{r}\right\Vert \leq1$, with the
origin representing a maximally mixed state. The angles $\theta$ and $\varphi$
are the polar and the azimuthal angles, respectively. (b) The Poincar\'{e}
sphere of unit radius for the polarized state of a beam of light. A point
$P=P\left(  \beta\text{, }\chi\right)  $ on the surface of the Poincar\'{e}
sphere is defined by the vector $\vec{s}\overset{\text{def}}{=}\left(
S_{1}\text{, }S_{2}\text{, }S_{3}\right)  =\left(  S_{0}\cos2\beta\cos
2\chi\text{, }S_{0}\cos2\beta\sin2\chi\text{, }S_{0}\sin2\beta\right)  $ where
$\vec{S}\overset{\text{def}}{=}\left(  S_{0}\text{, }S_{1}\text{, }%
S_{2}\text{, }S_{3}\right)  $ with $S_{1}^{2}+S_{2}^{2}+S_{3}^{2}=S_{0}%
^{2}\equiv1$ is the Stokes vector specified by the four Stokes parameters
$\left\{  S_{0}\text{, }S_{1}\text{, }S_{2}\text{, }S_{3}\right\}  $. The
parameter $S_{0}$ denotes the total intensity of the beam while the remaining
three parameters $S_{1}$, $S_{2}$ and, $S_{3}$ specify the polarization state
of the beam. A partially polarized beam of light is specified by $S_{0}\leq1$,
with the origin being a completely unpolarized beam. Finally, $\beta$ and
$\chi$ are the ellipticity and the orientation angles, respectively.}%
\end{figure}

\subsection{Coherence of the electric vibrations}

Having discussed the basics of polarized light, in this subsection we present
the essentials concerning the notion of coherence of the electric vibrations.

Given the electric vector of the incident light wave in its complex form, the
so-called coherency matrix $J$ is defined as \cite{wolf59},%
\begin{equation}
J=\left(
\begin{array}
[c]{cc}%
J_{xx} & J_{xy}\\
J_{yx} & J_{yy}%
\end{array}
\right)  \overset{\text{def}}{=}\left(
\begin{array}
[c]{cc}%
\left\langle E_{x}E_{x}^{\ast}\right\rangle  & \left\langle E_{x}E_{y}^{\ast
}\right\rangle \\
\left\langle E_{y}E_{x}^{\ast}\right\rangle  & \left\langle E_{y}E_{y}^{\ast
}\right\rangle
\end{array}
\right)  \text{,} \label{coherency}%
\end{equation}
where the angle brackets denote time average. The coherency matrix $J$ is an
Hermitian matrix with $J_{xy}^{\ast}=J_{yx}$ and characterizes the incident
wave. In particular, \textrm{tr}$\left(  J\right)  $ represents the intensity
of the incident wave and its off-diagonal coefficients describe the
correlation between the $x$- and $y$-components of $\vec{E}$. We observe that
employing the Schwarz inequality for integrals, it follows that $\left\vert
J_{xy}\right\vert \leq\sqrt{J_{xx}}\sqrt{J_{yy}}$ and $\left\vert
J_{yx}\right\vert \leq\sqrt{J_{yy}}\sqrt{J_{xx}}$. Therefore, $\det\left(
J\right)  \overset{\text{def}}{=}J_{xx}J_{yy}-J_{xy}J_{yx}\geq0$. The Stokes
parameters can be expressed in terms of the coherency matrix coefficients by
the relations $S_{0}\overset{\text{def}}{=}J_{xx}+J_{yy}$, $S_{1}%
\overset{\text{def}}{=}J_{xx}-J_{yy}$, $S_{2}\overset{\text{def}}{=}%
J_{xy}+J_{yx}$, and $S_{3}\overset{\text{def}}{=}i\left(  J_{yx}%
-J_{xy}\right)  $. Inverting these equations, one gets $J_{xx}=\left(
S_{0}+S_{1}\right)  /2$, $J_{yy}=\left(  S_{0}-S_{1}\right)  /2$,
$J_{xy}=\left(  S_{2}+iS_{3}\right)  /2$, and $J_{yx}=\left(  S_{2}%
-iS_{3}\right)  /2$. Therefore, the relation between the Stokes parameters
$\left\{  S_{0}\text{, }S_{1}\text{, }S_{2}\text{, }S_{3}\right\}  $ and the
coherency matrix $J$ can be recast in the following compact form
\cite{wolf59,fano54},%
\begin{equation}
J=\frac{1}{2}\sum_{i=0}^{3}S_{i}\sigma_{i}\text{,} \label{gege}%
\end{equation}
where in Eq. (\ref{gege}) $\sigma_{0}=I_{2\times2}$, $\sigma_{1}=\sigma_{z}$,
$\sigma_{2}=\sigma_{x}$, $\sigma_{3}=-\sigma_{y}$ with $\left\{  \sigma
_{x}\text{, }\sigma_{y}\text{, }\sigma_{z}\right\}  $ being the usual Pauli
spin matrices in quantum mechanics. To quantify the electric vibrations in the
$\hat{x}$- and $\hat{y}$-directions, we introduce the so-called complex degree
of coherence
\begin{equation}
j_{xy}=\left\vert j_{xy}\right\vert e^{i\beta_{xy}}\overset{\text{def}}%
{=}\frac{J_{xy}}{\sqrt{J_{xx}}\sqrt{J_{yy}}}\text{.} \label{coherence}%
\end{equation}
In Eq. (\ref{coherence}), $\left\vert j_{xy}\right\vert $ is the modulus of
the complex degree of coherence ( we shall call it, degree of coherence) and
measures the degree of correlation of the vibrations. The phase $\beta_{xy}\in%
\mathbb{R}
$, instead, specifies the effective phase difference between the vibrations.
As a side remark, we note that $\det\left(  J\right)  \geq0$ implies
$\left\vert j_{xy}\right\vert \leq1$. We notice that $J$ in Eq.
(\ref{coherency}) will change if the $\hat{x}$- and $\hat{y}$-axes are rotated
about the direction of propagation of the wave. Therefore, since $\left\vert
j_{xy}\right\vert $ is not expressed in terms of rotation-invariant terms, it
depends on the choice of the $\hat{x}$- and $\hat{y}$-axes. Unlike $\left\vert
j_{xy}\right\vert $, the degree of polarization \textrm{P }in Eq.
(\ref{polarization})\textrm{ }of a wave can be expressed in terms of
rotation-invariant quantities built from the coherency matrix as we shall see
in the next subsection.

\subsection{Degree of polarization and coherency matrix}

To express the degree of polarization of a wave in terms of the coherency
matrix, we proceed as follows. Recall that a general coherency matrix
$J_{\mathrm{general}}$ can be formally recast as,%
\begin{equation}
J_{\mathrm{general}}=\left(
\begin{array}
[c]{cc}%
J_{xx} & J_{xy}\\
J_{yx} & J_{yy}%
\end{array}
\right)  \overset{\text{def}}{=}\left(
\begin{array}
[c]{cc}%
\alpha_{1} & \gamma_{1}-i\delta_{1}\\
\gamma_{1}+i\delta_{1} & \beta_{1}%
\end{array}
\right)  \text{,} \label{general}%
\end{equation}
with $\alpha_{1}$, $\beta_{1}$, $\gamma_{1}$, $\delta_{1}\in%
\mathbb{R}
$. Moreover, recall that any wave can be represented as a superposition of a
wave of natural radiation with coherency matrix $J_{\mathrm{natural}}%
\overset{\text{def}}{=}\left[  D^{2}\text{, }0\text{; }0\text{, }D^{2}\right]
$ and a completely elliptically polarized (monochromatic) wave with coherency
matrix $J_{\mathrm{pol}}\overset{\text{def}}{=}\left[  A^{2}\text{,
}-iAB\text{; }iAB\text{, }B^{2}\right]  $ with $A$, $B$, $D\in%
\mathbb{R}
$. Then, it can be shown that all light is a case or limiting case of
partially elliptically polarized light with coherency matrix given by
$J_{\mathrm{tot}}\overset{\text{def}}{=}J_{\mathrm{natural}}+J_{\mathrm{pol}}%
$,%
\begin{equation}
J_{\mathrm{tot}}=\left(
\begin{array}
[c]{cc}%
A^{2}+D^{2} & -iAB\\
iAB & B^{2}+D^{2}%
\end{array}
\right)  \text{.} \label{total}%
\end{equation}
To prove this statement, it is sufficient to show there exists a
transformation that allows us to set Eq. (\ref{total}) equal to Eq.
(\ref{general}). Indeed, it turns out that $J_{\mathrm{tot}}=T\left(
\chi\right)  J_{\mathrm{general}}T^{-1}\left(  \chi\right)  $ where $T\left(
\chi\right)  \overset{\text{def}}{=}\left[  \cos\chi\text{, }\sin\chi\text{;
}\sin\chi\text{, }-\cos\chi\right]  $ is a real unitary transformation with
$\chi$ being the angle (that is, the orientation angle for the polarization
ellipse that corresponds to the light beam) defined by the condition
\cite{wiener30},%
\begin{equation}
\tan\left(  2\chi\right)  =\frac{2\gamma_{1}}{\alpha_{1}-\beta_{1}}%
=\frac{J_{xy}+J_{yx}}{J_{xx}-J_{yy}}\text{.} \label{impo1}%
\end{equation}
For further details on how to express $A$, $B$, and $D$ in terms of
$\alpha_{1}$, $\beta_{1}$, $\gamma_{1}$, $\delta_{1}$, we refer to Ref.
\cite{wiener30}. Now, setting $J_{xx}\overset{\text{def}}{=}A^{2}+D^{2}$,
$J_{xy}\overset{\text{def}}{=}-iAB$, $J_{yx}\overset{\text{def}}{=}iAB$, and
$J_{yy}\overset{\text{def}}{=}B^{2}+D^{2}$, we finally have%
\begin{equation}
\mathrm{P}\overset{\text{def}}{=}\frac{I_{\mathrm{pol}}}{I_{\mathrm{tot}}%
}=\frac{\mathrm{tr}\left(  J_{\mathrm{pol}}\right)  }{\mathrm{tr}\left(
J_{\mathrm{tot}}\right)  }=\left[  1-\frac{4\det\left(  J_{\mathrm{tot}%
}\right)  }{\left[  \mathrm{tr}\left(  J_{\mathrm{tot}}\right)  \right]  ^{2}%
}\right]  ^{1/2}\text{.} \label{pollo}%
\end{equation}
From Eq. (\ref{pollo}), we note that $\mathrm{P}$ does not depend on the
choice of the $\hat{x}$- and $\hat{y}$-directions. Furthermore, from Eq.
(\ref{pollo}) and the definition of $\left\vert j_{xy}\right\vert $, we obtain
after some algebra that $\left\vert j_{xy}\right\vert \leq\mathrm{P}$
\cite{wolf59}. The equality $\left\vert j_{xy}\right\vert =\mathrm{P}$ holds
iff $J_{xx}=J_{yy}$. It can be shown that a pair of orthogonal directions
$\hat{x}^{\prime}$ and $\hat{y}^{\prime}$ always exist for which this is the case.

The fact that $\left\vert j_{xy}\right\vert $ depends on the choice of the
$\hat{x}$ and $\hat{y}$ directions while $\mathrm{P}$ does not, along with the
definition of the angle $\chi$ in Eq. (\ref{impo1}), will play a major role in
our discussion of unit optical efficiency in the next section.

\section{Propagation of light with unit optical efficiency}

\begin{table}[t]
\centering
\begin{tabular}
[c]{c|c|c}\hline\hline
\textbf{Type of sphere} & \textbf{Constraint description} & \textbf{Constraint
equation}\\\hline
Bloch & Bounded energy of the system & $\left(  E_{+}-E_{-}\right)
^{2}=\left(  h_{11}-h_{22}\right)  ^{2}+4h_{12}h_{21}=\mathrm{fixed}$\\
Bloch & Maximal energy dispersion & $\Delta E=\Delta E_{\max}$, and $\Delta
t=\Delta t_{\min}$\\
Poincar\'{e} & Bounded intensity of light & $I_{\mathrm{pol}}^{2}%
=(J_{xx}-J_{yy})^{2}+4J_{xy}J_{yx}=\mathrm{fixed}$\\
Poincar\'{e} & Maximal correlations between $E_{x}$ and $E_{y}$ & $S_{2}%
^{2}=\left(  S_{2}^{2}\right)  _{\max}$, and $S_{1}^{2}=0$\\\hline
\end{tabular}
\caption{Schematic summary of the main constraint equations yielding unit
efficiency geodesic paths on the Bloch sphere and optical paths leading to
polarization states with maximal degree of coherence on the Poincar\'{e}
sphere.}%
\end{table}

In this section, we finally describe the propagation of polarized light with
maximal degree of coherence.

Let us define a measure of optical efficiency as the ratio between the degree
of polarization of the wave and the degree of coherence of the electric
vibrations, $\eta_{\mathrm{opt}}\overset{\text{def}}{=}\left\vert
j_{xy}\right\vert /\mathrm{P}$. This quantity achieves its maximum value $1$
when $\left\vert j_{xy}\right\vert =\mathrm{P}$, that is to say, when
$J_{xx}=J_{yy}$. For a fixed value of $\mathrm{P}$ or, analogously, for a
fixed value of $I_{\mathrm{pol}}\overset{\text{def}}{=}\mathrm{tr}\left(
J_{\mathrm{pol}}\right)  =\left[  \left(  J_{xx}+J_{yy}\right)  ^{2}%
+4\det\left(  J_{\mathrm{pol}}\right)  \right]  ^{1/2}=$\textrm{constant
}\cite{wolf59}, we wish to find a new pair of orthogonal directions $\left\{
\hat{x}^{\prime}\text{, }\hat{y}^{\prime}\right\}  $ such that $J_{x^{\prime
}x^{\prime}}=J_{y^{\prime}y^{\prime}}$ and, consequently, $\eta_{\mathrm{opt}%
}=1$. First, using the definition of $\det\left(  J_{\mathrm{tot}}\right)  $,
we note that the constraint on $I_{\mathrm{pol}}$ can be recast as%
\begin{equation}
I_{\mathrm{pol}}^{2}=\left(  J_{xx}-J_{yy}\right)  ^{2}+4J_{xy}J_{yx}%
=\mathrm{const}\text{.} \label{coco}%
\end{equation}
Therefore, from Eq. (\ref{coco}) we have that the optimal coherency matrix
$J^{\prime}$ is specified by $J_{x^{\prime}x^{\prime}}=J_{y^{\prime}y^{\prime
}}$ and $\left\vert J_{x^{\prime}y^{\prime}}\right\vert =I_{\mathrm{pol}}/2$.
Observe that Eq. (\ref{coco}) is the analog of Eq. (\ref{ec}). Furthermore,
the quantum conditions $h_{11}=h_{22}$ and $h_{12}^{\max}=E_{0}/2$ correspond
to the optical conditions $J_{x^{\prime}x^{\prime}}=J_{y^{\prime}y^{\prime}}$
and $\left\vert J_{x^{\prime}y^{\prime}}\right\vert =I_{\mathrm{pol}}/2$, respectively.

Alternatively, in terms of the Stokes vector components, unit optical
efficiency demands%
\begin{equation}
S_{1}^{2}\rightarrow\left(  S_{1}^{2}\right)  _{\min}=0\text{, with }S_{2}%
^{2}\rightarrow\left(  S_{2}^{2}\right)  _{\max}\text{.} \label{coco1}%
\end{equation}
Note that the minimization of $S_{1}^{2}$ and the maximization of $S_{2}^{2}$
correspond to the minimization of the evolution time and the maximization of
the energy uncertainty, respectively. In Table II, we present a schematic of
the main constraint equations yielding unit efficiency geodesic paths on the
Bloch sphere [Eqs. (\ref{ec}) and (\ref{ec2})] and optical paths leading to
polarization states with maximal degree of coherence on the Poincar\'{e}
sphere [Eqs. (\ref{coco}) and (\ref{coco1})].

To find the pair of orthogonal directions $\left\{  \hat{x}^{\prime}\text{,
}\hat{y}^{\prime}\right\}  $, we assume they are obtained from the canonical
Cartesian directions $\left\{  \hat{x}\text{, }\hat{y}\right\}  $ via a
rotation $R_{\hat{z}}\left(  \varphi_{\mathrm{opt}}\right)  \overset
{\text{def}}{=}\left[  \cos\varphi_{\mathrm{opt}}\text{, }\sin\varphi
_{\mathrm{opt}}\text{; }-\sin\varphi_{\mathrm{opt}}\text{, }\cos
\varphi_{\mathrm{opt}}\right]  $ around the $\hat{z}$-axis by an angle
$\varphi_{\mathrm{opt}}$ to be determined. Specifically, the components of the
electric field $\vec{E}$ with respect to the basis $\left\{  \hat{x}^{\prime
}\text{, }\hat{y}^{\prime}\right\}  $ satisfy%
\begin{equation}
\left[  \vec{E}\right]  _{\left\{  \hat{x}\text{, }\hat{y}\right\}
}\rightarrow\left[  \vec{E}\right]  _{\left\{  \hat{x}^{\prime}\text{, }%
\hat{y}^{\prime}\right\}  }\overset{\text{def}}{=}R_{\hat{z}}\left(
\varphi_{\mathrm{opt}}\right)  \cdot\left[  \vec{E}\right]  _{\left\{  \hat
{x}\text{, }\hat{y}\right\}  }\text{.} \label{et}%
\end{equation}
The transformation laws for the coherency matrix and the Stokes vector
emerging from Eq. (\ref{et}) are given by \cite{kim19},
\begin{equation}
J\rightarrow J^{\prime}\overset{\text{def}}{=}R_{\hat{z}}\left(
\varphi_{\mathrm{opt}}\right)  \cdot J\cdot R_{\hat{z}}\left(  -\varphi
_{\mathrm{opt}}\right)  \text{,} \label{jj}%
\end{equation}
and \cite{chipman19},%
\begin{equation}
S\rightarrow S^{\prime}\overset{\text{def}}{=}M_{\mathrm{ROT}}\left(
\varphi_{\mathrm{opt}}\right)  S=\mathcal{U}\cdot\left[  R_{\hat{z}}^{\ast
}\left(  \varphi_{\mathrm{opt}}\right)  \otimes R_{\hat{z}}\left(
\varphi_{\mathrm{opt}}\right)  \right]  \cdot\mathcal{U}^{\dagger}\text{,}
\label{geo2}%
\end{equation}
respectively, where $\mathcal{U}$ is a $\left(  4\times4\right)  $-unitary
matrix and $M_{\mathrm{ROT}}\left(  \varphi_{\mathrm{opt}}\right)  $ is a
$\left(  4\times4\right)  $- Mueller matrix given by%
\begin{equation}
\mathcal{U}\overset{\text{def}}{=}\frac{1}{\sqrt{2}}\left(
\begin{array}
[c]{cccc}%
1 & 0 & 0 & 1\\
1 & 0 & 0 & -1\\
0 & 1 & 1 & 0\\
0 & i & -i & 0
\end{array}
\right)  \text{, and }M_{\mathrm{ROT}}\left(  \varphi_{\mathrm{opt}}\right)
\overset{\text{def}}{=}\left(
\begin{array}
[c]{cccc}%
1 & 0 & 0 & 0\\
0 & \cos\left(  2\varphi_{\mathrm{opt}}\right)  & \sin\left(  2\varphi
_{\mathrm{opt}}\right)  & 0\\
0 & -\sin\left(  2\varphi_{\mathrm{opt}}\right)  & \cos\left(  2\varphi
_{\mathrm{opt}}\right)  & 0\\
0 & 0 & 0 & 1
\end{array}
\right)  \text{,} \label{rot}%
\end{equation}
respectively. Imposing that $J_{x^{\prime}x^{\prime}}=J_{y^{\prime}y^{\prime}%
}$, from Eq. (\ref{jj}) we get that $\varphi_{\mathrm{opt}}$ is such that%
\begin{equation}
\tan\left(  2\varphi_{\mathrm{opt}}\right)  =\frac{J_{yy}-J_{xx}}%
{J_{xy}+J_{yx}}\text{.} \label{impo2}%
\end{equation}
Since $J_{xx}$, $J_{yy}$, and $J_{xy}+J_{yx}=2\operatorname{Re}\left(
J_{xy}\right)  \in%
\mathbb{R}
$, Eq. (\ref{impo2}) has a real root. In conclusion, there always exists a
pair of orthogonal directions $\left\{  \hat{x}^{\prime}\text{, }\hat
{y}^{\prime}\right\}  $ with $\hat{x}^{\prime}\overset{\text{def}}{=}\hat
{x}\cos\varphi_{\mathrm{opt}}+\hat{y}\sin\varphi_{\mathrm{opt}}$ and $\hat
{y}^{\prime}\overset{\text{def}}{=}-\hat{x}\sin\varphi_{\mathrm{opt}}+\hat
{y}\cos\varphi_{\mathrm{opt}}$ for which the two intensities $J_{xx}$ and
$J_{yy}$ are equal. For this pair of directions, the degree of coherence
$\left\vert j_{xy}\right\vert $ reaches its maximum value $\left\vert
j_{xy}\right\vert _{\max}$ with $\left\vert j_{xy}\right\vert _{\max
}=\mathrm{P}$. This particular pair of directions $\left\{  \hat{x}^{\prime
}\text{, }\hat{y}^{\prime}\right\}  $ has a neat geometric interpretation.
Indeed, using Eqs. (\ref{impo1}) and (\ref{impo2}), it follows that%
\begin{equation}
\tan\left(  2\varphi_{\mathrm{opt}}\right)  \tan\left(  2\chi\right)
=-1\text{,} \label{impo2A}%
\end{equation}
that is, $\varphi_{\mathrm{opt}}-\chi=\pi/4$ or $3\pi/4$. Therefore, the
directions $\left\{  \hat{x}^{\prime}\text{, }\hat{y}^{\prime}\right\}  $ for
which $\eta_{\mathrm{opt}}\overset{\text{def}}{=}\left\vert j_{xy}\right\vert
/\mathrm{P}=1$ are the bisectors of the principal directions $\left\{
\hat{\xi}\text{, }\hat{\eta}\right\}  $ with $\hat{\xi}\overset{\text{def}}%
{=}\hat{x}\cos\chi+\hat{y}\sin\chi$ and $\hat{\eta}\overset{\text{def}}%
{=}-\hat{x}\sin\chi+\hat{y}\cos\chi$ of the polarization ellipse of the
polarized portion of the wave \cite{wolf59}. Therefore, given that $\hat
{x}^{\prime}=\hat{\xi}\cos\left(  \varphi_{\mathrm{opt}}-\chi\right)
+\hat{\eta}\sin\left(  \varphi_{\mathrm{opt}}-\chi\right)  $ and $\hat
{y}^{\prime}=-\hat{\xi}\sin\left(  \varphi_{\mathrm{opt}}-\chi\right)
+\hat{\eta}\cos\left(  \varphi_{\mathrm{opt}}-\chi\right)  $ with
$\varphi_{\mathrm{opt}}-\chi=\pi/4$ or $3\pi/4$, we have%
\begin{equation}
\left\vert \hat{x}^{\prime}\cdot\hat{\xi}\right\vert =\left\vert \hat
{x}^{\prime}\cdot\hat{\eta}\right\vert =\left\vert \hat{y}^{\prime}\cdot
\hat{\xi}\right\vert =\left\vert \hat{y}^{\prime}\cdot\hat{\eta}\right\vert
=1/2\text{,} \label{you1}%
\end{equation}
since $\hat{x}^{\prime}=\left(  \hat{\xi}+\hat{\eta}\right)  /\sqrt{2}$ and
$\hat{y}^{\prime}=\left(  \hat{\eta}-\hat{\xi}\right)  /\sqrt{2}$. Observe
that the optical conditions in Eq. (\ref{you1}) correspond to the quantum
conditions
\begin{equation}
\left\vert \left\langle E_{+}|A\right\rangle \right\vert =\left\vert
\left\langle E_{-}|A\right\rangle \right\vert =\left\vert \left\langle
E_{+}|A_{\bot}\right\rangle \right\vert =\left\vert \left\langle E_{-}%
|A_{\bot}\right\rangle \right\vert =1/2\text{,} \label{you2}%
\end{equation}
obtained when maximizing the energy uncertainty in Eq. (\ref{chi4}). In Table
III, we provide a schematic description of quantities of interest on the Bloch
and the Poincar\'{e} spheres. In particular, we characterize the points
$P\left(  \theta\text{, }\varphi\right)  $ and $P\left(  \beta\text{, }%
\chi\right)  $ on the two surfaces in terms of their spherical coordinates.
Moreover, we specify the rotation operations yielding unit efficiency on the
two spheres by means of their axes of rotation (that is, $\hat{E}_{+}$ and
$\hat{z}$, respectively) and their angles of rotation (that is, $\cos
^{-1}\left[  \left\vert \left\langle A|B\right\rangle \right\vert \right]  $
and $\varphi_{\mathrm{opt}}$, respectively). Finally, we determine the two
angles $\varphi_{E_{+}}$ and $\varphi_{\mathrm{opt}}$ to be compared within
the two geometric frameworks of unit efficiency quantum evolutions and unit
efficiency polarized light propagation.

In the next section, we discuss the physical root that is underlying our
proposed formal analogy.\begin{table}[t]
\centering
\begin{tabular}
[c]{c|c|c}\hline\hline
\textbf{Quantity of interest} & \textbf{Bloch sphere} & \textbf{Poincar\'{e}
sphere}\\\hline
Angles & $\left(  \theta\text{, }\varphi\right)  $ & $\left(  \beta\text{,
}\chi\right)  $\\
Range of angles & $0\leq\theta\leq\pi$, $0\leq\varphi<2\pi$ & $-\pi
/4<\beta\leq\pi/4$, $0\leq\chi<\pi$\\
Point on the sphere & $P\left(  \theta\text{, }\varphi\right)  \overset
{\text{def}}{=}\left(  \sin\theta\cos\varphi\text{, }\sin\theta\sin
\varphi\text{, }\cos\theta\right)  $ & $P\left(  \beta\text{, }\chi\right)
\overset{\text{def}}{=}\left(  \cos2\beta\cos2\chi\text{, }\cos2\beta\sin
2\chi\text{, }\sin2\beta\right)  $\\
Axis of rotation & $\hat{E}_{+}=\hat{E}_{+}\left(  \theta_{E_{+}}\text{,
}\varphi_{E_{+}}\right)  $ & $\hat{z}$, fixed\\
Angle of rotation & $\cos^{-1}\left[  \left\vert \left\langle A|B\right\rangle
\right\vert \right]  $, fixed & $\varphi_{\mathrm{opt}}$\\
Angles to be compared & $\varphi_{E_{+}}=\varphi_{E_{+}}\left(  \left\vert
A\right\rangle \text{, }\left\vert B\right\rangle \right)  $ & $\varphi
_{\mathrm{opt}}=\varphi_{\mathrm{opt}}\left(  P\right)  $, with $P=P\left(
\beta\text{, }\chi\right)  $\\\hline
\end{tabular}
\caption{Schematic description of quantities of interest on the Bloch and the
Poincar\'{e} spheres. In particular, we compare the description of points
$P\left(  \theta\text{, }\varphi\right)  $ and $P\left(  \beta\text{, }%
\chi\right)  $ on the two surfaces in terms of their spherical coordinates.
Moreover, we describe the rotation operations yielding unit efficiency on the
two spheres in terms of their axes of rotation (that is, $\hat{E}_{+}$ and
$\hat{z}$, respectively) and their angles of rotation (that is, $\cos
^{-1}\left[  \left\vert \left\langle A|B\right\rangle \right\vert \right]  $
and $\varphi_{\mathrm{opt}}$, respectively). Finally, we identify the two
angles $\varphi_{E_{+}}$ and $\varphi_{\mathrm{opt}}$ to be compared within
the two geometric frameworks of unit efficiency quantum evolutions and
polarization optics.}%
\end{table}

\section{Physical origin behind the formal analogy}

The formal analogy between quantum evolutions with unit geometric efficiency
and propagation of light with unit optical efficiency, discussed in this paper
and summarized in Tables I and II, is yet another example of the close
relationship that exists between certain classes of quantum mechanical and
classical optical phenomena. Is this analogy completely unexpected? What is
the physical reason that underlines such a formal similarity? This analogy is
not completely unexpected.\ After all, as mentioned in the Introduction,
Grover exploited his knowledge on the interference of classical light waves in
order to construct his quantum search algorithm. Furthermore, the set of
unit-speed quantum mechanical evolutions includes as a special case the
Farhi-Gutmann search Hamiltonian \cite{farhi98}, an analog version of Grover's
digital quantum search scheme. Both Grover's and the Farhi-Gutmann search
schemes rely heavily on the interference phenomenon for achieving their
quadratic speedup. The phenomenon of interference, either constructive or
destructive, plays a key role in both light propagation
\cite{wiener30,mandel91} and quantum searching \cite{cleve98A,cleve98B}. For
additional details on the role played by interference effects in light
propagation, quantum searching, and optimal-speed quantum evolutions, we refer
to Appendix D.

In what follows, we briefly discuss the role played by interference as the
physical root underlying the formal analogy between propagation of light with
maximal degree of coherence and optimal-speed unitary quantum propagation
proposed in this paper.

\subsection{Interference of classical light waves}

In the framework of coherent light propagation \cite{wiener30}, two rays of
light originating from the same source can interfere. Specifically, the two
rays can be combined in such a manner to give rise to a light more intense
than is ordinarily created by two light beams of their respective intensities
(constructive interference). Alternatively, the superimposition of the two
rays of light can yield a darkness (destructive interference). Therefore,
coherent light propagation is characterized by interference effects where, in
addition, the degree of coherence is equal to the degree of
indistinguishability of the particle trajectories that yield the interference
pattern \cite{mandel91}. When the photon pattern becomes identifiable, the
interference effects disappear, and the light propagation becomes incoherent.
In the study of coherence properties of partially polarized electromagnetic
radiation \cite{wolf59}, there is a proper angle that specifies a pair of
directions for which the degree of coherence of the electric vibrations has
its maximum value (which, in turn, equals the degree of polarization of the
wave). As discussed in this paper, this angle $\varphi_{\mathrm{opt}}$ is
determined by a specific value of the orientation angle $\chi$ that
characterizes the polarization ellipse used to describe the light propagation
[see Eq. (\ref{impo2A})].

From a more quantitative standpoint, consider a quasi-monochromatic light wave
that propagates in the $\hat{z}$-direction specified by an electric field
$\vec{E}\left(  t\right)  =E_{x}\left(  t\right)  \hat{x}+E_{y}\left(
t\right)  \hat{y}$. Assume that the component $E_{\theta}=\vec{E}\cdot
\hat{\theta}$ of the electric field in the $\hat{\theta}$-direction is given
by, $E_{\theta}\left(  t\text{; }\theta\text{, }\varepsilon\right)
\overset{\text{def}}{=}E_{x}\left(  t\right)  \cos\left(  \theta\right)
+E_{y}e^{i\varepsilon}\sin\left(  \theta\right)  $, with $\varepsilon$
denoting the phase delay between $E_{x}$ and $E_{y}$. The interference law of
light waves can be expressed by calculating the intensity $I\left(
\theta\text{, }\varepsilon\right)  $ of the light vibrations in the direction
which makes an angle $\theta$ with the positive $\hat{x}$-direction. A
straightforward calculation yields \cite{wolf59},%
\begin{equation}
I\left(  \theta\text{, }\varepsilon\right)  =I_{x}+I_{y}+2\sqrt{I_{x}}%
\sqrt{I_{y}}\left\vert j_{xy}\right\vert \cos\left(  \beta_{xy}-\varepsilon
\right)  \text{. } \label{intensityA}%
\end{equation}
In Eq. (\ref{intensityA}), $I\left(  \theta\text{, }\varepsilon\right)
\overset{\text{def}}{=}\left\langle E_{\theta}\left(  t\text{; }\theta\text{,
}\varepsilon\right)  E_{\theta}^{\ast}\left(  t\text{; }\theta\text{,
}\varepsilon\right)  \right\rangle $ where angle brackets denote time average,
$I_{x}\overset{\text{def}}{=}J_{xx}\cos^{2}\left(  \theta\right)  $,
$I_{y}\overset{\text{def}}{=}J_{yy}\sin^{2}\left(  \theta\right)  $, $J_{ij}$
are the coefficients of the coherency matrix, and $j_{xy}\overset{\text{def}%
}{=}\left\vert j_{xy}\right\vert e^{i\beta_{xy}}$ is the complex degree of
coherence of the electric vibrations in the $\hat{x}$- and $\hat{y}%
$-directions. Recall that $\left\vert j_{xy}\right\vert $ in Eq.
(\ref{intensityA}) is an indicator of the degree of correlation of the
vibrations, while $\beta_{xy}$ in Eq. (\ref{intensityA}) is an effective phase
difference between the electric vibrations in the $\hat{x}$- and $\hat{y}%
$-directions. In modern terminology, we emphasize that $J_{ij}$ are known as
the coefficients of the polarization matrix \cite{wolf07}. Moreover, in the
context of the classical theory of optical fluctuations and coherence, the
analog of $\left\vert j_{xy}\right\vert $ is the so-called degree of first
order coherence \cite{loudon00}. Regardless of notation and modern
terminology, what is most important for us here is the contribution of
$j_{xy}$ with the interference term $\left\vert j_{xy}\right\vert \cos\left(
\beta_{xy}-\varepsilon\right)  $ into the expression of the total intensity
$I\left(  \theta\text{, }\varepsilon\right)  $ in Eq. (\ref{intensityA}).

Interestingly, when studying the superposition of two coherent beams of light
in different states of elliptic polarization, Pancharatnam showed in Ref.
\cite{pancharatnam56} that if $A$ and $B$ represent the states of polarization
on the Poincar\'{e} sphere of the given interfering beams, and $C$ that of the
resultant beam, the intensity of the resultant beam can be recast as%
\begin{equation}
I_{C}=I_{A}+I_{B}+2\sqrt{I_{A}}\sqrt{I_{B}}\cos\left(  \frac{\theta
_{AB}^{\mathrm{Poincar\acute{e}}}}{2}\right)  \cos\left(  \delta\right)
\text{.} \label{intensityB}%
\end{equation}
In Eq. (\ref{intensityB}), $\theta_{AB}^{\mathrm{Poincar\acute{e}}}$ is the
angular separation of states $A$ and $B$ on the Poincar\'{e} sphere, while
$\delta$ is not quite the absolute difference of phase between the two beams
and is defined as the phase advance of the first beam being in a state of
polarization $A$ over the $A$-component of the second beam being in a state of
polarization $B$. If we set $\varepsilon=\varepsilon_{y}$ in Eq.
(\ref{intensityA}) and consider a nonvanishing phase $\varepsilon_{x}$,
$\delta$ can be formally identified with $\varepsilon_{x}-\left(
\varepsilon_{y}-\beta_{xy}\right)  $. For more details, we refer to Ref.
\cite{pancharatnam56}.

\subsection{Interference of quantum probability amplitudes}

In the framework of quantum searching viewed in the context of quantum
computing as multi-particle interference \cite{cleve98A,cleve98B}, the role of
interference is fundamental since it permits the evolution from a source state
to a target state by manipulating the intermediate multi-particle
superpositions in a convenient way. Specifically, quantum searching can be
regarded as inducing a proper relative phase between two eigenvectors to
generate constructive interference on the searched elements and destructive
interference on the remaining ones. As pointed out in this paper, this phase
is quantified by a specific value of the azimuthal angle $\varphi_{E_{+}}$
that specifies the location on the Bloch sphere of the eigenstates $\left\vert
E_{\pm}\right\rangle $ used to geometrically construct the optimal evolution
(search) Hamiltonian \textrm{H}.

Therefore, interference appears to be the essential physical phenomenon that
underlies both propagation of light with maximal degree of coherence and
continuous-time quantum search evolution with minimum search time (i.e.,
optimal-speed). Interference is optimally exploited in the two above mentioned
tasks so that unit optical and quantum efficiencies can be achieved by
identifying suitable angles. These are the orientation and azimuthal angles in
the optical and quantum search cases, respectively. The orientation angle
$\varphi_{\mathrm{opt}}$ specifies the optimal unitary operation [Mueller
rotation, $M_{\mathrm{ROT}}\left(  \varphi_{\mathrm{opt}}\right)  $] that
connects the two initial and final polarization states on the Poincar\'{e}
sphere. The azimuthal angle $\varphi_{E_{+}}$, instead, characterizes the
optimal unitary operation [Bloch rotation, $R_{\hat{E}_{+}}\left(  \theta
_{AB}\right)  $ with $\theta_{AB}\overset{\text{def}}{=}\cos^{-1}\left[
\left\vert \left\langle A|B\right\rangle \right\vert \right]  $ and $\hat
{E}_{+}\overset{\text{def}}{=}\hat{E}_{+}\left(  \theta_{E_{+}}\text{,
}\varphi_{E_{+}}\right)  $] that connects the source and the target states
$\left\vert A\right\rangle $ and $\left\vert B\right\rangle $ on the Bloch sphere.

As mentioned earlier, the essential prerequisite for achieving speedups in
quantum searching is interference of quantum probability amplitudes
\cite{ekert96,galindo02}. This occurs in both Grover's original quantum search
algorithm \cite{grover97} and in the Farhi-Gutmann continuous version of
Grover's algorithm \cite{farhi98}. Indeed, as pointed out by Lloyd in Ref.
\cite{lloyd99}, Grover arrived at the formulation of his quantum search
algorithm inspired by the interference of classical waves emitted by an array
of antennas.

From an explicit viewpoint, consider a quantum state $\left\vert
\psi\right\rangle $ written as the superposition of two normalized quantum
states $\left\vert A\right\rangle $ and $\left\vert B\right\rangle $ with
complex probability amplitudes $a\overset{\text{def}}{=}\left\vert
a\right\vert e^{i\varphi_{a}}$ and $b\overset{\text{def}}{=}\left\vert
b\right\vert e^{i\varphi_{b}}$, respectively, with $\varphi_{a}$ and
$\varphi_{b}$ in $%
\mathbb{R}
$. Furthermore, let us assume that $\left\langle A|B\right\rangle
=\left\langle B|A\right\rangle ^{\ast}\overset{\text{def}}{=}\left\vert
\left\langle A|B\right\rangle \right\vert e^{i\varphi_{AB}}$ with
$\varphi_{AB}\in%
\mathbb{R}
$. Then, the interference law of probability amplitudes $a$ and $b$ can be
expressed in terms of their corresponding probabilities calculated by taking a
modulus squared of the probability amplitudes. After some simple algebra in
which we consider the inner product of $\left\vert \psi\right\rangle $ with
itself, the quantum interference law becomes%
\begin{equation}
p_{a+b}=p_{a}+p_{b}+2\sqrt{p_{a}}\sqrt{p_{b}}\left\vert \left\langle
A|B\right\rangle \right\vert \cos\left[  \varphi_{AB}-\left(  \varphi
_{a}-\varphi_{b}\right)  \right]  \text{,} \label{intensityC}%
\end{equation}
where $p_{a+b}\overset{\text{def}}{=}\left\langle \psi|\psi\right\rangle $,
$p_{a}\overset{\text{def}}{=}\left\vert a\right\vert ^{2}$, and $p_{b}%
\overset{\text{def}}{=}\left\vert b\right\vert ^{2}$. In Eq. (\ref{intensityC}%
), $\left\vert \left\langle A|B\right\rangle \right\vert $ can be viewed as
$\cos\left(  \theta_{AB}^{\mathrm{Bloch}}/2\right)  $ with $\theta
_{AB}^{\mathrm{Bloch}}$ being the geodesic distance on the Bloch sphere
between the states $\left\vert A\right\rangle $ and $\left\vert B\right\rangle
$, while $\varphi_{a}-\varphi_{b}$ is the absolute phase difference between
the interfering probability amplitudes $a$ and $b$. Interestingly, we remark
the contribution of $\left\langle A|B\right\rangle \overset{\text{def}}%
{=}\left\vert \left\langle A|B\right\rangle \right\vert e^{i\varphi_{AB}}$
with the interference term $\left\vert \left\langle A|B\right\rangle
\right\vert \cos\left[  \varphi_{AB}-\left(  \varphi_{a}-\varphi_{b}\right)
\right]  $ into the expression of the total probability $p_{a+b}$ in Eq.
(\ref{intensityC}).

Considering Eqs. (\ref{intensityA}), (\ref{intensityB}), and (\ref{intensityC}%
), we note that $\left\langle A|B\right\rangle \overset{\text{def}}%
{=}\left\vert \left\langle A|B\right\rangle \right\vert e^{i\varphi_{AB}}$
corresponds to $j_{xy}\overset{\text{def}}{=}\left\vert j_{xy}\right\vert
e^{i\beta_{xy}}$. In particular, we get%
\begin{equation}
\frac{\left\vert J_{xy}\right\vert }{\sqrt{J_{xx}}\sqrt{J_{yy}}}%
\overset{\text{def}}{=}\left\vert j_{xy}\right\vert \leftrightarrow\cos\left(
\frac{\theta_{AB}^{\mathrm{Poincar\acute{e}}}}{2}\right)  \leftrightarrow
\cos\left(  \frac{\theta_{AB}^{\mathrm{Bloch}}}{2}\right)  \overset
{\text{def}}{=}\frac{\left\vert \left\langle A|B\right\rangle \right\vert
}{\sqrt{\left\langle A|A\right\rangle }\sqrt{\left\langle B|B\right\rangle }%
}\text{.} \label{analogies}%
\end{equation}
Equation (\ref{analogies}) is especially relevant since the degree of
correlation of the electric vibrations ($\left\vert j_{xy}\right\vert $)
viewed in terms of the angular separation on the Poincar\'{e} sphere
($\theta_{AB}^{\mathrm{Poincar\acute{e}}}$) can be regarded as corresponding
to the geodesic distance on the Bloch sphere ($\theta_{AB}^{\mathrm{Bloch}}$).
In the analysis carried out in our paper, both $\cos\left(  \theta
_{AB}^{\mathrm{Bloch}}/2\right)  $ and $\left\vert j_{xy}\right\vert $ played
a major role in the proposed definitions of quantum geometric efficiency
$\eta_{\mathrm{QM}}\overset{\text{def}}{=}s_{0}/s$ and classical optical
efficiency $\eta_{\mathrm{opt}}\overset{\text{def}}{=}\left\vert
j_{xy}\right\vert /\mathrm{P}$.

In summary, we have discussed the link between propagation of light with
maximal degree of coherence and optimal-speed quantum propagation by
performing a punctual comparative analysis of geometric flavor. The emergence
of this formal analogy is physically motivated by the existence of a key
physical phenomenon that underlies both types of evolutions at their best,
i.e., interference. This link among $\left\vert j_{xy}\right\vert $,
$\theta_{AB}^{\mathrm{Poincar\acute{e}}}$, and $\theta_{AB}^{\mathrm{Bloch}}$
in\ Eq. (\ref{analogies}) that emerges while thinking of interference of
classical waves (classical optics) and interference of probability amplitudes
(quantum mechanics) should be kept in mind as a constant (hidden) theme
underlying our discussion in the main paper.

\section{Concluding remarks}

We present here a summary of our main findings along with a discussion on
possible future applications of our work.

\subsection{Summary of results}

In this paper, we identified and discussed\textbf{ }in a quantitative manner a
link between the geometry of time-independent optimal-speed Hamiltonian
quantum evolutions on the Bloch sphere and the geometry of
intensity-preserving propagation of light with maximal degree of coherence on
the Poincar\'{e} sphere.

Specifically, we carried out a detailed comparative analysis between the
quantum and optical scenarios. In the quantum case, we focused on the main
constraint equations [Eqs. (\ref{ec}) and (\ref{ec2})] leading to the
construction of the optimal unitary evolution operator $e^{-\frac{i}{\hslash
}\mathrm{H}T_{AB}}$ (that is, a rotation of a Bloch vector on the Bloch
sphere) with the optimal Hamiltonian given in Eqs. (\ref{hopt}) and
(\ref{amy}). In the optical case, similarly, we focused on the main constraint
equations [Eqs. (\ref{coco}) and (\ref{coco1})] leading to the construction of
the optimal Mueller matrix $M_{\mathrm{ROT}}\left(  \varphi_{\mathrm{opt}%
}\right)  $ [see Eq. (\ref{rot}) with $\varphi_{\mathrm{opt}}$
in\ Eq.(\ref{impo2})]. This Mueller matrix acts on a Stokes vector on the
Poincar\'{e} sphere [see Eq. (\ref{geo2})] and leads to the propagation of
light with maximal degree of coherence in analogy to the geodesic path defined
in\ Eq. (\ref{geodesic}) and generated by the Hamiltonian in Eq. (\ref{amy}).
In particular, in Table I we presented an explicit correspondence between the
main quantum and optical quantities that enter the two phenomena. In Table II,
we pointed out the two main constraint relations that specify the two physical
scenarios. Finally, in Table III, we concluded with the correspondence between
axes and angles of rotations that specify the two optimal operations yielding
unit quantum geometric efficiency and classical optical efficiency, respectively.

Our main achievement in this paper is bringing to light this fascinating
analogy between optimal-speed quantum evolutions and polarized light
propagation with maximal degree of coherence. This link constitutes a
different connection between the quantum physics of two-level systems and
classical polarization optics. To a certain extent, we think that our
investigation is not only relevant from a pure theoretical perspective, it can
also be regarded (in retrospect) as providing a sort of conceptual and
quantitative geometric background underlying Grover's powerful intuition about
constructing a quantum search scheme by mimicking interference of classical
waves \cite{lloyd99}.

Clearly, it could be worthwhile exploring the possibility of extending our
work to higher-dimensional quantum systems since a single two-level quantum
system is so simple that connecting it to classical wave propagation may not
necessarily make the two-level system more intuitive. A richer Hilbert space
structure would be more appropriate to fully gain physical insights emerging
from our proposed analogy. Therefore, we expect that it would be very helpful
outlining the needed formalism to generalize our current result to multi-qubit
quantum systems and demonstrate, for instance, that classical optics is an
intuitive way to understand entangled quantum systems. This is a crucial step
that we leave to future scientific efforts since it goes beyond the scope of
the paper. However, in the next subsection, we do explore in a qualitative
manner some possible future line of investigations that emerge from our analysis.

\subsection{Outlook}

In addition to its intrinsic conceptual and pedagogical values, our
theoretical study paves the way to several intriguing explorative physics
questions that require more attention.

First, regarding the time-optimal Hamiltonian analysis within the general
setting specified by the so-called quantum brachistochrone problem (QBP,
\cite{carlini06,campaioli19}), it may be of interest investigating
similarities between propagation of light with maximal degree of coherence and
the QBP. In this context, one may think of extending our investigation to
higher-dimensional spin systems \cite{fry08} and to constraint equations
specifying cost functions other than time-optimality
\cite{russell14,russell15}. In our paper, we have provided the optical analog
of a QBP via its analogy with the propagation of light with maximal degree of
coherence. The QBP was characterized by a cost functional defined in terms of
time-optimality to be optimized by imposing a single constraint, a bound on
the energy resource. In general, defining the cost functional, an efficiency
measure of getting to a target state from a given initial state by a suitable
choice of a Hamiltonian, is a nontrivial task that depends on the physical
scenario being investigated. As pointed out in the Introduction, there is a
variety of cost functionals that one may consider in QBPs and, in addition,
the optimization procedure may by specified by multiple constraints to be
simultaneously satisfied. A typical set of examples includes minimization of
the total amount of time \cite{carlini06}, the heating rate \cite{deffner14},
the energy dispersion rate \cite{deffner14}, and the entropy production rate
\cite{cafaropre20,gassner21}. It would be certainly interesting from a physics
perspective uncovering possible optical analogs of such more general QBPs. We
think that identifying suitable optical constraints would be a key step in
this direction and one may need to go beyond considering constraints expressed
in terms of the intensity of light.

Second, being in the framework of quantum speed limit (QSL) problems
concerning the minimum time needed to transfer a given initial quantum state
to a final one \cite{mande45,margo98,deffner17}, it appears that changes in
coherence have significant dynamical effects on the evolution speeds of the
reduced state of certain families of quantum gases viewed as interacting
many-particle systems \cite{xu20}. More specifically, in Ref. \cite{xu20} the
authors study the dynamics of the reduced single-particle density matrix
(RSPDM) of a strongly correlated bosonic quantum gas in one dimension and a
gas of spinless fermions. They focus on two dynamical processes, a sudden
quench, and the efficient control of the system by means of a shortcut to
adiabaticity. The physics of the gases is characterized in terms of the
time-averaged Schatten-1 norm of the dynamics, that is the speed of evolution
of the system. Furthermore, the coherence of the gases is specified by means
of the largest eigenvalue of the RSPDM, a good measure of the presence of
off-diagonal long-range order. The authors state in Ref. \cite{xu20} that
coherences play an essential role in the evolution of the reduced state of the
systems. In the case of strongly interacting bosons, they find larger average
speeds (thus, smaller quantum speed limit times) due to the presence of
off-diagonal excitations emerging from the scattering between particles. In
our work, the minimum quantum speed limit time is achieved in the presence of
maximal energy dispersion with the speed of evolution of the quantum system
being proportional to the energy dispersion of the Hamiltonian operator.
Furthermore, our results suggest that maximal energy dispersion corresponds
from an optical standpoint to maximal correlations between the orthogonal
electric field components of the light wave that appear as off-diagonal terms
in the coherency matrix. Thus, the minimum quantum speed limit time appears to
correspond to a maximal correlational structure in the field components
specifying the electromagnetic radiation. Given these formal similarities
between our work and the one in Ref. \cite{xu20}, it seems rather intriguing
exploring if our analysis might help further understanding the role played by
coherence in the control of many-body quantum states. We believe that a first
significant step in this direction would be exploring the possible existence
of a quantitative connection between the degree of coherence employed in our
work and the coherence specified by means of the largest eigenvalue of the RSPDM.

Third, when studying quantum resources, it happens that the quantum Fisher
information and the super-radiant quantity attributed to coherence are
antithetical resources. Specifically, there is a trade-off between the quantum
Fisher information and the super-radiant quantity \cite{tan18}. The trade-off
emerges in a coherence limited scenario where optimizing one quantity seems to
suggest less quantum resources that can be utilized for the other.
Interestingly, identifying the energy uncertainty and the degree of coherence
with the quantum Fisher information and the super-radiant quantity,
respectively, we also find there appears to be a conflicting behavior between
these two quantities. Indeed, considering the unit efficiency scenario where
time-optimality is the resource to be optimized, to an increase of one of
these two quantities there corresponds necessarily a decrease of the other one
for a fixed minimum total amount of time for the evolution. To further
elaborate on this point, we remark that in Ref. \cite{rossatto20} the authors
study the Dicke model of superradiance specified by a system of $N$ identical
two-level atoms with transition frequency $\omega$ and interacting in a
collective fashion with the surrounding electromagnetic field in the vacuum
state at zero temperature. They find that the $l_{1}$ norm of coherence of the
single-atom density operator is proportional to the square root of the
normalized average radiation intensity emitted in a cooperative manner by the
whole superradiant system. This radiation intensity, in turn, can be recast in
terms of the coherence of the normalized total electric dipole moment of the
system. Thus, an important link between the $l_{1}$ norm of coherence of the
single-atom density operator and the coherence of the normalized total
electric dipole moment of the system is emphasized. This link leads to the
validation of the $l_{1}$ norm of coherence as a figure of merit of the
superradiance phenomenon in the mean-field approach. As a main finding, the
authors showed that the evolution of the system is faster when more coherence
is stored in the single-atom state. Interestingly, our investigation also
leads to the conclusion that optimal evolution speed corresponds to maximal
degree of coherence. Our work could be potentially relevant for better
understanding the reason why quantum coherence speeds up the evolution of
superradiant systems. We anticipate that a basic preliminary step in this
direction would be that of clarifying the relation between the degree of
coherence employed in our work and the $l_{1}$ norm of coherence, a very
intuitive and easy to use coherence measure related to off-diagonal elements
of a quantum state with the key feature of being the most general coherence
monotone introduced in Ref. \cite{plenio14} and discussed with emphasis on its
applications in Ref. \cite{plenio17}.

Fourth, it is pointed out in Ref. \cite{plenio17} that quantum coherence can
also be used as a resource in quantum algorithms. For instance, it is
emphasized that the success probability in the analog Grover algorithm depends
on the amount of coherence, quantified via the $l_{1}$ norm of coherence, in
the corresponding quantum state \cite{pati16,fan17}. Uncovering possible links
between our current work and the findings presented in Refs.
\cite{pati16,fan17} could be yet another intriguing avenue to explore in
future investigations. Given the physically intuitive link with off-diagonal
elements of both the $l_{1}$ norm of coherence and the degree of coherence
used in our work, we remark once again that a much needed step in this
direction would be investigating the possibility of quantifying light
propagation by means of the $l_{1}$ norm of coherence.

We hope our work will inspire other scientists and pave the way toward further
investigations in this fascinating research direction. For the time being, we
leave a more in-depth quantitative discussion on these potential extensions
and applications of our theoretical findings to quantum brachistochrone
problems, quantum speed limits questions, and quantum resources analyses to
future scientific efforts.\bigskip

\begin{acknowledgments}
C.C. is grateful to the United States Air Force Research Laboratory (AFRL)
Summer Faculty Fellowship Program for providing support for this work. S.R.
acknowledges support from the National Research Council Research Associate
Fellowship program (NRC-RAP). P.M.A. acknowledges support from the Air Force
Office of Scientific Research (AFOSR). Any opinions, findings and conclusions
or recommendations expressed in this material are those of the author(s) and
do not necessarily reflect the views of the Air Force Research Laboratory
(AFRL). The authors thank the anonymous referees for stimulating comments
leading to an improved version of the manuscript.
\end{acknowledgments}

\appendix

\section{Mueller matrices}

In this appendix, we provide further details on the Mueller matrices\textbf{
}mentioned in Sec. III B. Furthermore, we devote special emphasis on their
relation with the Jones matrices. Finally, we emphasize how points on the
surface of the Poincar\'{e} sphere are rotated by means of specific types of
Mueller matrices.

In the traditional approach to polarization optics, the light propagates along
the $\hat{z}$-axis and one considers the electric vector field components
along the $\hat{x}$- \ and $\hat{y}$-directions. The polarization state is
determined by the amplitude ratio and phase difference of the electric field
components. Therefore, polarization can be modified either by changing the
amplitudes or by tuning the relative phases, or both. Within the Jones
calculus \cite{jones41,jones47}, the Jones vector in $%
\mathbb{C}
^{2}$ represents the polarized light by means of the amplitude and the phase
of the electric field in the $\hat{x}$- \ and $\hat{y}$-directions.
Furthermore, linear optical elements (for instance, beam splitters, lenses,
and mirrors) are represented by $2\times2$ Jones matrices. Within the Mueller
calculus \cite{mueller48}, employing the concepts of Stokes parameters and
Poincar\'{e} sphere, the change of polarization due to the interaction of
light with an optical device can be described by the action of a $4\times4$
matrix that represents a linear transformation acting upon a $4\times1$ matrix
corresponding to the Stokes vector. Within the Mueller calculus, there are
three fundamental optical elements: wave plates, rotators, and polarizers. A
wave plate and a rotator produce phase shifts and rotations of the Stokes
vector, respectively. They are described by unitary matrices since they do not
change the intensity of the light. Polarizers cause anisotropic attenuation
and do change the intensity of light passing through them. Therefore, unlike
wave plates and rotators, they are described by nonunitary matrices. The
matricial representation of optical devices is very useful. Indeed, the
composite effect of a series of optical devices crossed by a light beam is
represented by the product of the matrices corresponding to the various
optical elements in the series. Mueller matrices can be grouped into two main
categories \cite{goldberg20}: nondepolarizing and depolarizing Mueller
matrices. Nondepolarizing Mueller matrices can modify the degree of
polarization of partially polarized light. However, they do not change the
degree of polarization of perfectly polarized light. Depolarizing Mueller
matrices, instead, while maintaining the total intensity of the light beam, do
reduce the degree of polarization of completely polarized light. Furthermore,
nondepolarizing Mueller matrices have equivalent Jones matrices. On the other
hand, depolarizing Mueller matrices have no equivalent Jones matrices. For a
discussion on necessary and sufficient conditions for a Mueller matrix to be
derivable from a Jones matrix, we refer to Ref. \cite{anderson94}.
Interestingly, it is possible to show that any Mueller matrix can be
decomposed into a sequence of three matrix factors \cite{lu96}: a
diattenuator, followed by a retarder, then followed by a depolarizer.
Diattenuators and retarders are described by Hermitian Jones matrices and
change only the amplitudes of the components of the electric field vector. A
polarizer is an example of a diattenuator. Retarders, instead, are described
by unitary Jones matrices and change only the phases of components of the
electric field vector. A wave plate is an example of a retarder. As mentioned
earlier, there are Mueller matrices with no corresponding Jones matrices.
However, it turns out that any Jones matrix $J$ acting on the electric field
$\vec{E}$ can be transformed into the corresponding Mueller matrix $M$ given
by $M\overset{\text{def}}{=}A\left(  J\otimes J^{\ast}\right)  A^{-1}$, where
\textquotedblleft$\ast$\textquotedblright\ and \textquotedblleft$\otimes
$\textquotedblright\ denote the complex conjugate and the tensor product,
respectively. Moreover, $A$ is a $4\times4$ matrix defined as,%
\begin{equation}
A\overset{\text{def}}{=}\left(
\begin{array}
[c]{cccc}%
1 & 0 & 0 & 1\\
1 & 0 & 0 & -1\\
0 & 1 & 1 & 0\\
0 & -i & i & 0
\end{array}
\right)  \text{.} \label{AAA}%
\end{equation}
Observe that the four rows $\left\{  R_{A}^{i}\right\}  _{1\leq i\leq4}$ of
$A$ in Eq. (\ref{AAA}) are given by the coefficients of the identity matrix
$I$ and the three Pauli matrices $\left\{  \sigma_{x}\text{, }\sigma
_{y}\text{, }\sigma_{z}\right\}  $ with $R_{A}^{1}\leftrightarrow I$,
$R_{A}^{2}\leftrightarrow\sigma_{z}$, $R_{A}^{3}\leftrightarrow\sigma_{x}$,
and $R_{A}^{4}\leftrightarrow\sigma_{y}$. \ It is well-known that there is a
two-to-one homomorphism between the complex special unitary group $SU(2)$ and
the real group of three-dimensional pure rotations $O^{+}\left(  3\right)  $
\cite{ryder96},%
\begin{equation}
SU(2)\ni e^{i\vec{\sigma}\cdot\hat{n}\frac{\theta}{2}}\leftrightarrow
e^{i\vec{J}\cdot\hat{n}\theta}\in O^{+}\left(  3\right)  \text{.} \label{homo}%
\end{equation}
In Eq. (\ref{homo}), $\hat{n}$ denotes the axis of rotation, $\theta$ is the
angle of rotation, $\vec{\sigma}\overset{\text{def}}{=}\left(  \sigma
_{1}\text{, }\sigma_{2}\text{, }\sigma_{3}\right)  $ is the Pauli matrix
vector, and $\vec{J}\overset{\text{def}}{=}\left(  J_{1}\text{, }J_{2}\text{,
}J_{3}\right)  $ is the generator vector for $O^{+}\left(  3\right)  $.
Interestingly, it can be shown that the matrix coefficients $R_{ij}$ of any
rotation matrix $R$ in $O^{+}\left(  3\right)  $ can be recast as
\cite{wigner59},%
\begin{equation}
R_{ij}=\frac{1}{2}\mathrm{tr}\left(  U^{\dagger}\cdot\sigma_{i}\cdot
U\cdot\sigma_{j}\right)  \text{,} \label{homo2}%
\end{equation}
where $U$ is a two-dimensional unitary matrix with determinant equal to one
specified by three free (real) parameters. From Eqs. (\ref{homo}) and
(\ref{homo2}), we note there is a global topological difference between
$SU(2)$ and $O^{+}\left(  3\right)  $. For example, increasing the angle
$\theta$ by $2\pi$ in Eq. (\ref{homo}), we get $U\rightarrow-U$ in $SU(2)$
while $R\rightarrow R$ in $O^{+}\left(  3\right)  $. Therefore, both $U$ and
$-U$ in $SU(2)$ correspond to the same $R$ in $O^{+}\left(  3\right)  $. Thus,
there exists a two-to-one mapping of elements of $SU(2)$ onto $O^{+}\left(
3\right)  $. Exploiting the $SU\left(  2\right)  $-$O^{+}\left(  3\right)  $
homomorphism, it happens that to a $SU\left(  2\right)  $ matrix $U$ acting on
the vector field $\vec{E}$ there corresponds a $4\times4$ Mueller matrix
viewed as an augmented form of a $O^{+}\left(  3\right)  $ matrix $R$
\cite{cloude86},
\begin{equation}
M\overset{\text{def}}{=}\left(
\begin{array}
[c]{cc}%
1_{1\times1} & O_{1\times3}\\
O_{3\times1} & R_{3\times3}%
\end{array}
\right)  \text{,}%
\end{equation}
where the coefficients $M_{ij}$ of the matrix $M$ are given by,%
\begin{equation}
M_{ij}\overset{\text{def}}{=}\frac{1}{2}\mathrm{tr}\left(  U^{\dagger}\cdot
\Xi_{i}\cdot U\cdot\Xi_{j}\right)  \text{,} \label{remark}%
\end{equation}
with $\vec{\Xi}\overset{\text{def}}{=}\left(  I\text{, }\sigma_{z}\text{,
}\sigma_{x}\text{, }\sigma_{y}\right)  $. For completeness, we remark that Eq.
(\ref{remark}) can also be extended by considering arbitrary complex
(scattering) matrices $U_{\operatorname{complex}}$ in place of $SU\left(
2\right)  $ matrices. In this case, the effect of the Mueller matrix with
coefficients $M_{ij}\overset{\text{def}}{=}1/2\mathrm{tr}\left(
U_{\operatorname{complex}}^{\dagger}\cdot\Xi_{i}\cdot
U_{\operatorname{complex}}\cdot\Xi_{j}\right)  $ is to combine a rotation of
the Stokes vector with a change of its length (which, in turn, corresponds to
a change in the degree of polarization). For more details on polarization
algebra with Mueller matrices, we refer to Ref. \cite{cloude86}.

\section{Degree of coherence of partially polarized waves}

In this appendix, we use the Poincar\'{e} sphere formalism to describe the
behavior of the modulus $\left\vert j_{xy}\right\vert $ of the complex degree
of coherence $j_{xy}$ of partially polarized light beams in terms of the
ellipticity and orientation angles. This appendix helps better understanding
of the content of Sec. III B with regard to partially polarized light waves
with \textrm{P}$<1$\textbf{.}

Partially polarized waves can be regarded as points that are inside the
Poincar\'{e} sphere of radius $I_{\mathrm{tot}}$ (i.e., the total intensity of
the wave) and at a distance $I_{\mathrm{pol}}$ (i.e., the intensity of the
polarized part of the wave) from the origin of the sphere itself. Using the
Stokes parameters $\left\{  S_{0}\text{, }S_{1}\text{, }S_{2}\text{, }%
S_{3}\right\}  $, we note that the degree of polarization \textrm{P }and
$\left\vert j_{xy}\right\vert $ can be recast as,%
\begin{equation}
\mathrm{P}=\frac{\left(  S_{1}^{2}+S_{2}^{2}+S_{3}^{2}\right)  ^{1/2}}{S_{0}%
}\text{, and }\left\vert j_{xy}\right\vert =\left(  \frac{S_{2}^{2}+S_{3}^{2}%
}{S_{0}^{2}-S_{1}^{2}}\right)  ^{1/2}\text{,} \label{fifi1}%
\end{equation}
respectively. From Eq. (\ref{fifi1}), we have that if $S_{0}^{2}=S_{1}%
^{2}+S_{2}^{2}+S_{3}^{2}$ then \textrm{P}$=\left\vert j_{xy}\right\vert =1$.
Instead, if $S_{0}^{2}>S_{1}^{2}+S_{2}^{2}+S_{3}^{2}$ then \textrm{P}$<1$ and
$\left\vert j_{xy}\right\vert \leq$\textrm{P}. Furthermore, using the two
relations in Eq. (\ref{fifi1}) along with setting $S_{0}\overset{\text{def}%
}{=}I_{\mathrm{tot}}$, $S_{1}\overset{\text{def}}{=}I_{\mathrm{pol}}%
\cos\left(  2\beta\right)  \cos\left(  2\chi\right)  $, $S_{2}\overset
{\text{def}}{=}I_{\mathrm{pol}}\cos\left(  2\beta\right)  \sin\left(
2\chi\right)  $, and $S_{3}\overset{\text{def}}{=}I_{\mathrm{pol}}\sin\left(
2\beta\right)  $, we get%
\begin{equation}
\left\vert j_{xy}\right\vert =\left\vert j_{xy}\right\vert \left(
\beta\text{, }\chi\text{; }\mathrm{P}\right)  \overset{\text{def}}%
{=}\mathrm{P}\left[  \frac{1-\cos^{2}\left(  2\beta\right)  \cos^{2}\left(
2\chi\right)  }{1-\mathrm{P}^{2}\cos^{2}\left(  2\beta\right)  \cos^{2}\left(
2\chi\right)  }\right]  ^{1/2}\text{,} \label{fifi}%
\end{equation}
where $\mathrm{P}\overset{\text{def}}{=}I_{\mathrm{pol}}/I_{\mathrm{tot}}$,
$-\pi/4<\beta\leq\pi/4$, and $0\leq\chi<\pi$. Finally, we note from Eq.
(\ref{fifi}) that for a given value of $\mathrm{P}<1$ and for any value of the
ellipticity angle $\beta\in\left(  -\pi/4\text{, }\pi/4\right]  $, the maximum
of $\left\vert j_{xy}\right\vert $ equals $\mathrm{P}$ and is achieved when
the orientation angle $\chi$ equals $\pi/4$.

\section{Parametrizations of qubits and polarization states}

In this appendix, we report for completeness some mathematical details on the
parametrization of qubits and polarization states viewed as points on the
Bloch sphere and the Poincar\'{e} sphere, respectively. These details help
comprehend the schematic depictions in Fig. $1$.

In terms of the computational basis vectors $\left\vert 0\right\rangle $ and
$\left\vert 1\right\rangle $, a normalized qubit is a point on the Bloch
sphere that can be parametrized as%
\begin{equation}
\left\vert \psi\left(  \theta\text{, }\varphi\right)  \right\rangle
\overset{\text{def}}{=}\cos\left(  \frac{\theta}{2}\right)  \left\vert
0\right\rangle +e^{i\varphi}\sin\left(  \frac{\theta}{2}\right)  \left\vert
1\right\rangle \text{,} \label{Q1}%
\end{equation}
with $0\leq\theta\leq\pi$ and $0\leq\varphi<2\pi$. The Bloch sphere metric is
given by $ds_{\mathrm{Bloch}}^{2}\overset{\text{def}}{=}d\hat{n}_{\mathrm{B}%
}\cdot d\hat{n}_{\mathrm{B}}=d\theta^{2}+\sin^{2}\left(  \theta\right)
d\varphi^{2}$ with $\hat{n}_{\mathrm{B}}\overset{\text{def}}{=}\left\langle
\psi\left(  \theta\text{, }\varphi\right)  |\vec{\sigma}|\psi\left(
\theta\text{, }\varphi\right)  \right\rangle =\left(  \sin\theta\cos
\varphi\text{, }\sin\theta\sin\varphi\text{, }\cos\theta\right)  $ and
$\vec{\sigma}\overset{\text{def}}{=}\left(  \sigma_{x}\text{, }\sigma
_{y}\text{, }\sigma_{z}\right)  $ being the usual vector of Pauli matrices.

In terms of the orthonormal circular basis states $\hat{e}_{\mathrm{RC}}$
(right circular) and $\hat{e}_{\mathrm{LC}}$ (left circular), the general
equation of a normalized state of polarization $\hat{e}\left(  \beta\text{,
}\chi\right)  $ viewed as a point on the Poincar\'{e} sphere is given by%
\begin{equation}
\hat{e}\left(  \beta\text{, }\chi\right)  \overset{\text{def}}{=}\frac
{\cos\left(  \beta\right)  +\sin\left(  \beta\right)  }{\sqrt{2}}\hat
{e}_{\mathrm{RC}}+e^{i2\chi}\frac{\cos\left(  \beta\right)  -\sin\left(
\beta\right)  }{\sqrt{2}}\hat{e}_{\mathrm{LC}}\text{,} \label{Q2}%
\end{equation}
with $-\pi/4<\beta\leq\pi/4$ and $0\leq\chi<\pi$. The Poincar\'{e} metric is
given by $ds_{\mathrm{Poincar\acute{e}}}^{2}\overset{\text{def}}{=}d\hat
{n}_{\mathrm{P}}\cdot d\hat{n}_{\mathrm{P}}=4\left[  d\beta^{2}+\cos
^{2}\left(  2\beta\right)  d\chi^{2}\right]  $ where $\hat{n}_{\mathrm{P}%
}\overset{\text{def}}{=}\left\langle \hat{e}\left(  \beta\text{, }\chi\right)
\text{, }\vec{\sigma}\hat{e}\left(  \beta\text{, }\chi\right)  \right\rangle
_{%
\mathbb{C}
}=\left(  \cos\left(  2\beta\right)  \cos\left(  2\chi\right)  \text{, }%
\cos\left(  2\beta\right)  \sin\left(  2\chi\right)  \text{, }\sin\left(
2\beta\right)  \right)  $, with $\left\langle \cdot\text{, }\cdot\right\rangle
_{%
\mathbb{C}
}$ denoting the usual complex inner product. From Eq. (\ref{Q2}), note that
points on the poles specify circularly polarized light, $\hat{e}_{\mathrm{RC}%
}\overset{\text{def}}{=}\hat{e}\left(  \pi/4\text{, }0\right)  $ and $\hat
{e}_{\mathrm{LC}}\overset{\text{def}}{=}\hat{e}\left(  -\pi/4\text{,
}0\right)  $. Furthermore, points on the equator correspond to linearly
polarized light, $\hat{e}_{\mathrm{VL}}\overset{\text{def}}{=}\hat{e}\left(
0\text{, }\pi/2\right)  $ (vertical linear) and $\hat{e}_{\mathrm{HL}}%
\overset{\text{def}}{=}\hat{e}\left(  0\text{, }0\right)  $ (horizontal
linear). Ignoring an overall phase, we remark that $\hat{e}_{\mathrm{RC}%
}\overset{\text{def}}{=}\left(  \hat{e}_{\mathrm{HL}}-i\hat{e}_{\mathrm{VL}%
}\right)  /\sqrt{2}$ and $\hat{e}_{\mathrm{LC}}\overset{\text{def}}{=}\left(
\hat{e}_{\mathrm{HL}}+i\hat{e}_{\mathrm{VL}}\right)  /\sqrt{2}$. Finally, the
remaining points on the Poincar\'{e} sphere represent other elliptical
polarization states.

\section{Interference effects in propagation of light, quantum searching, and
optimal-speed quantum evolutions}

In this appendix, we present some comments on the role played by interference
effects in light propagation, quantum searching, and optimal-speed quantum
evolutions. These remarks integrate those presented in Sec. V.

In the study of propagation of light, coherent sources are required for
producing interference patterns. In particular, interference of light beams
appear in the calculation of the total intensity of the resultant beam
obtained in terms of a superposition of two coherent beams. In this context,
the objective is to obtain propagation of light with maximal degree of
coherence. Maximization of the absolute value of the complex degree of
coherence, interpreted as a measure of the degree of correlation of the
electric vibrations of the wave, yield more visible interference patterns.
Indeed, the degree of coherence establishes in a formal manner how distinctly
visible is the interference pattern \cite{qureshi19}. We note that to have a
nonzero degree of coherence, the coherency matrix has to have nonvanishing
off-diagonal terms \cite{born50,wolf59}. This observation becomes especially
interesting when we recall that quantum computation derives its power from
entanglement and quantum interference. In particular, the degree of
interference in a $N$-qubit register is specified by the coherences, that is,
the off-diagonal elements $\rho_{lm}$ with $l\neq m$ of the density operator
in the computation basis. Therefore, the role played by coherences in
quantifying the degree of quantum interference viewed as a source of power for
quantum computing can be grasped in a straightforward manner \cite{ekert96}.
These considerations lead us to the following question: Where does the
phenomenon of quantum interference manifest itself in the quantum
computational tasks considered in our paper?

In Grover's digital quantum search algorithm \cite{grover97}, the interference
of quantum probability amplitudes appears in the calculation of the transition
probability from the (known) source state to the (unknown) target state.
Indeed, in quantum searching, the goal is to achieve unit transition
probability (defined as the modulus squared of the quantum overlap between the
target state and the source state acted upon by a number of iterations of
Grover's operator) with the smallest number of iterations of Grover's
operator. More specifically, quantum searching can be explained as inducing a
desired relative phase between two eigenvectors of Grover's operator to yield
constructive interference on the target state (that is, use quantum
interference to nudge up the searched state) and destructive interference on
the remaining states \cite{cleve98A,cleve98B}. In the Farhi-Gutmann analog
quantum search evolution viewed as the continuous-time version of Grover's
search scheme \cite{farhi98}, the interference of quantum probability
amplitudes emerges in the computation of the transition probability from the
(known) source state to the (unknown) target state. The goal there is to
achieve unit transition probability in the shortest amount of time. The
Farhi-Gutmann search Hamiltonian specifies the dynamical process of quantum
interference which, in turn, allows one to evolve from the source state to the
target state in the smallest possible time by modifying the explored
intermediate superpositions of quantum states in a suitably prescribed manner
so that time optimality \cite{carlini06,wang15} is achieved. Finally, in the
optimal-speed quantum Hamiltonian evolutions \cite{bender07,ali09},
interference of quantum probability amplitudes can be identified in the
maximization of the energy uncertainty (that is, the dispersion of the
Hamiltonian operator). This maximization is required in order to evolve from a
(known) source state to a (known) target state with optimal-speed (that is,
the maximum energy uncertainty and the smallest travel time). The
optimal-speed time-independent Hamiltonian specifies, via its eigenvector
decomposition, the process of quantum interference. The latter, in turn,
allows us to transition from the source to the target states in the shortest
possible time by navigating through a path of quantum states while preserving
maximal energy uncertainty. Interestingly, we pointed out the correspondence
between intensity of light and energy of the quantum system in the constraint
equations that appear in Table II.

\begin{thebibliography}{99}                                                                                               %


\bibitem {caneva09}T. Caneva, M. Murphy, T. Calarco, R. Fazio, S. Montagero,
V. Giovannetti, and G. E. Santoro, \emph{Optimal control at the quantum speed
limit}, Phys. Rev. Lett. \textbf{103}, 240501 (2009).

\bibitem {carlini13}V. Mukherjee, A. Carlini, A. Mari, T. Caneva, S.
Montagero, T. Calarco, R. Fazio, and V. Giovannetti,\emph{ Speeding up and
slowing down the relaxation of a qubit by optimal control}, Phys. Rev.
\textbf{A88}, 062326 (2013).

\bibitem {heger13}G. C. Hegerfeldt, \emph{Driving at the quantum speed limit:
Optimal control of a two-level system}, Phys. Rev. Lett. \textbf{111}, 260501 (2013).

\bibitem {boscain21}U. Boscain, M. Sigalotti, and D. Sugny, \emph{Introduction
to the Pontryagin maximum principle for quantum optimal control}, PRX Quantum
\textbf{2}, 030203 (2021).

\bibitem {brody15}D. C. Brody, G. W. Gibbons, and D. M. Meier,
\emph{Time-optimal navigation through quantum wind}, New J. Phys. \textbf{17},
033048 (2015).

\bibitem {deffner14}S. Deffner, \emph{Optimal control of a qubit in an optimal
cavity}, J. Phys. B: At. Mol. Opt. Phys. \textbf{47}, 145502 (2014).

\bibitem {fano54}U. Fano, \emph{A Stokes-parameter technique for the treatment
of polarization in quantum mechanics}, Phys. Rev. \textbf{93}, 121 (1954).

\bibitem {nielsen}M. A. Nielsen and I. L. Chuang, \emph{Quantum Computation
and Quantum Information}, Cambridge University Press (2000).

\bibitem {born50}M. Born and E. Wolf, \emph{Principles of Optics}, Cambridge
University Press (2003).

\bibitem {pancharatnam56}S. Pancharatnam, \emph{Generalized theory of
interference, and its applications}, Proc. Ind. Acad. Sci. \textbf{A44}, 247 (1956).

\bibitem {berry84}M. V. Berry, \emph{Quantal phase factors accompanying
adiabatic changes}, Proc. R. Soc. Lond. \textbf{A392}, 45 (1984).

\bibitem {berry87}M. V. Berry, \emph{The adiabatic phase and Pancharatnam's
phase for polarized light}, J. Mod. Opt. \textbf{34}, 1401 (1987).

\bibitem {samuel88}J. Samuel and R. Bhandari, \emph{General setting for
Berry's phase}, Phys. Rev. Lett. \textbf{60}, 2339 (1988).

\bibitem {wu86}R. Y. Chiao and Y.-S. Wu, \emph{Manifestation of Berry's
topological phase for the photon}, Phys. Rev. Lett. \textbf{57}, 933 (1986).

\bibitem {jordan87}T. F.\ Jordan, \emph{Direct calculation of the Berry phase
for spins and helicities}, J. Math. Phys. \textbf{28}, 1759 (1987).

\bibitem {cessa08}H. Moya-Cessa, J. R. Moya-Cessa, J. E. A. Landgrave, G.
Martinez-Niconoff, A. Perez-Leija, and A. T. Friberg, \emph{Degree of
polarization and quantum-mechanical purity}, J. Eur. Opt. Soc. Rap. Public.
\textbf{3}, 08014 (2008).

\bibitem {james12}O. Gamel and D. F. V. James, \emph{Measures of quantum state
purity and classical degree of polarization}, Phys. Rev. \textbf{A86}, 033830 (2012).

\bibitem {bender07}C. M. Bender, D. C. Brody, H. F. Jones, and B. K. Meister,
\emph{Faster than Hermitian quantum mechanics}, Phys. Rev. Lett. \textbf{98},
040403 (2007).

\bibitem {ali09}A. Mostafazadeh, \emph{Hamiltonians generating optimal-speed
evolutions}, Phys. Rev. \textbf{A79}, 014101 (2009).

\bibitem {brody06}D. C. Brody and D. W. Hook, \emph{On optimum Hamiltonians
for state transformations}, J. Phys. A: Math. Gen. \textbf{39}, L167 (2006).

\bibitem {brody07}D. C. Brody and D. W. Hook, \emph{On optimum Hamiltonians
for state transformation}, J. Phys. A: Math. Theor. \textbf{40}, 10949 (2007).

\bibitem {brody03}D. C. Brody, \emph{Elementary derivation for passage times},
J. Phys. A: Math. Gen. \textbf{36}, 5587 (2003).

\bibitem {anandan90}J. Anandan and Y. Aharonov, \emph{Geometry of quantum
evolution}, Phys. Rev. Lett. \textbf{65}, 1697 (1990).

\bibitem {cafaro20A}C. Cafaro, S. Ray, and P. M. Alsing, \emph{Geometric
aspects of analog quantum search evolutions}, Phys. Rev. \textbf{A102, }052607 (2020).

\bibitem {wolf59}E. Wolf, \emph{Coherence properties of partially polarized
electromagnetic radiation}, Il Nuovo Cimento \textbf{13}, 1180 (1959).

\bibitem {cafaro17}C. Cafaro, \emph{Geometric algebra and information geometry
for quantum computational software}, Physica \textbf{A470}, 154 (2017).

\bibitem {cafaro19A}C. Cafaro and P. M. Alsing, \emph{Theoretical analysis of
a nearly optimal analog quantum search}, Physica Scripta \textbf{94}, 085103 (2019).

\bibitem {cafaro19B}C. Cafaro and P. M. Alsing, \emph{Continuous-time quantum
search and time-dependent two-level quantum systems}, Int. J. Quantum
Information \textbf{17}, 1950025 (2019).

\bibitem {gassner20}S. Gassner, C. Cafaro, and S. Capozziello,
\emph{Transition probabilities in generalized quantum search Hamiltonian
evolutions}, Int. J. Geom. Meth. Mod. Phys. \textbf{17}, 2050006 (2020).

\bibitem {cafaro20B}C. Cafaro, D. Felice, and P. M. Alsing, \emph{Quantum
Groverian geodesic paths with gravitational and thermal analogies}, Eur. Phys.
J. Plus \textbf{135}, 900 (2020).

\bibitem {grover97}L. K. Grover, \emph{Quantum mechanics helps in searching
for a needle in a haystack}, Phys. Rev. Lett. \textbf{79}, 325 (1997).

\bibitem {lloyd99}S. Lloyd, \emph{Quantum search without entanglement}, Phys.
Rev. \textbf{A61}, 010301(R) (1999).

\bibitem {correction1}Note that our Eq. (\ref{hopt}) fixes a typographical
error that appears in Eq. (6) of Ref. \cite{bender07}. As a cross check, we
also verified that our Hamiltonian in Eq. (\ref{hopt}) satisfies the correct
maximal energy dispersion relation $\Delta E=$ $\Delta E_{\mathrm{\max}%
}\overset{\text{def}}{=}\left(  E_{+}-E_{-}\right)  /2=E_{0}/2$ with
$E_{+}-E_{-}\overset{\text{def}}{=}E_{0}=$\textrm{fixed}.

\bibitem {correction2}Observe that our Eq. (\ref{Hami}) fixes a typographical
error that appears in Eq. (12) of Ref. \cite{ali09}. As a cross check, we also
verified that our Hamiltonian in Eq. (\ref{Hami}) satisfies the correct
maximal energy dispersion relation given by\textbf{ }$\Delta E=$ $\Delta
E_{\mathrm{\max}}\overset{\text{def}}{=}\left(  E_{+}-E_{-}\right)  /2=E$ with
$E_{+}=-E_{-}\overset{\text{def}}{=}E$.

\bibitem {chandra50}S. Chandrasekhar, \emph{Radiative Transfer}, Clarendon
Press (1950).

\bibitem {collett68}E. Collett, \emph{The description of polarization in
classical physics}, Am. J. Phys. \textbf{36}, 713 (1968).

\bibitem {walker54}M. J.\ Walker, \emph{Matrix calculus and the Stokes
parameters of polarized radiation}, Am. J. Phys. \textbf{22}, 170 (1954).

\bibitem {jones41}R. C. Jones,\ \emph{A new calculus for the treatment of
optical systems. I. Description and discussion of the calculus}, J. Opt. Soc.
Am. \textbf{31}, 488 (1941); R. C. Jones, \emph{A new calculus for the
treatment of optical systems. IV}, J. Opt. Soc. Am. \textbf{32}, 486 (1942).

\bibitem {mueller48}H. Mueller, \emph{The foundation of optics}, J. Opt. Soc.
Am. \textbf{38}, 661 (1948).

\bibitem {stokes52}G. Stokes, \emph{On the composition and resolution of
streams of polarized light from different sources}, Trans. Cambridge
Phil.\ Soc. \textbf{9,} 399 (1852).

\bibitem {perrin42}F. Perrin,\emph{\ Polarization of light scattered by
isotropic opalescent media}, J. Chem. Phys. \textbf{10}, 415 (1942).

\bibitem {poincare92}H. Poincar\'{e}, \emph{Trait\'{e} de la Lumi\`{e}re},
Paris \textbf{2}, 165 (1892).

\bibitem {wiener30}N. Wiener, \emph{Generalized harmonic analysis},\ Acta
Math. \textbf{55}, 117 (1930).

\bibitem {kim19}S. Baskal, Y. S. Kim, and M. E. Noz, \emph{Mathematical
Devices for Optical Sciences}, IOP Publishing Ltd (2019).

\bibitem {chipman19}R. A. Chipman, W.-S. Tiffany Lam, and G.
Young,\emph{\ Polarized Light and Optical Systems}, CRC\ Press (2019).

\bibitem {farhi98}E. Farhi and S. Gutmann, \emph{Analog analogue of a digital
quantum computation}, Phys. Rev. \textbf{A57}, 2403 (1998).

\bibitem {mandel91}L. Mandel,\emph{\ Coherence and indistinguishability},
Optics Letters \textbf{16}, 1882 (1991).

\bibitem {cleve98A}R. Cleve, A. Ekert, L. Henderson, C. Macchiavello, and M.
Mosca, \emph{On quantum algorithms}, Complexity \textbf{4}, 33 (1998).

\bibitem {cleve98B}R. Cleve, A. Ekert, C. Macchiavello, and M. Mosca,
\emph{Quantum algorithms revisited}, Proc. Roy. Soc. Lond. \textbf{A454}, 339 (1998).

\bibitem {wolf07}E. Wolf,\emph{ Introduction to the Theory of Coherence and
Polarization of Light}, Cambridge University Press (2007).

\bibitem {loudon00}R. Loudon, \emph{The Quantum Theory of Light}, Oxford
University Press (2000).

\bibitem {ekert96}A. Ekert and R. Jozsa, \emph{Quantum computation and Shor's
factoring algorithm}, Rev. Mod. Phys. \textbf{68}, 733 (1996).

\bibitem {galindo02}A. Galindo and M. A. Martin-Delgado,\emph{ Information and
computation: classical and quantum aspects}, Rev. Mod. Phys. \textbf{74}, 347 (2002).

\bibitem {carlini06}A. Carlini, A. Hosoya, T. Koike, and Y. Okudaira,
\emph{Time-optimal quantum evolution}, Phys. Rev. Lett. \textbf{96}, 060503 (2006).

\bibitem {campaioli19}F. Campaioli, W. Sloan, K. Modi, and F. A. Pollock,
\emph{Algorithm for solving unconstrained unitary quantum brachistochrone
problems}, Phys. Rev. \textbf{A100}, 062328 (2019).

\bibitem {fry08}A. M. Frydryszak and V. M. Tkachuk, \emph{Quantum
brachistochrone problem for a spin-1 system in a magnetic field}, Phys. Rev.
\textbf{A77}, 014103 (2008).

\bibitem {russell14}B. Russell and S. Stepney, \emph{Zarmelo navigation and a
speed limit to quantum information processing}, Phys. Rev. \textbf{A90},
012303 (2014).

\bibitem {russell15}B. Russell and S. Stepney, \emph{Zarmelo navigation in the
quantum brachistochrone}, J. Phys. A: Math. Theor. \textbf{48}, 115303 (2015).

\bibitem {cafaropre20}C. Cafaro and P. M.\ Alsing, \emph{Information geometry
aspects of minimum entropy production paths from quantum mechanical
evolutions}, Phys. Rev. \textbf{E101}, 022110 (2020).

\bibitem {gassner21}S. Gassner, C. Cafaro, S. A. Ali, and P. M. Alsing,
\emph{Information geometric aspects of probability paths with minimum entropy
production for quantum state evolution}, Int. Geom. Meth. Mod. Phys.
\textbf{18}, 2150127 (2021).

\bibitem {mande45}L. Mandelstam and Ig. Tamm, \emph{The uncertainty relation
between energy and time in non-relativistic quantum mechanics}, J. Phys. USSR
\textbf{9}, 249 (1945).

\bibitem {margo98}N. Margolus and L. B. Levitin, \emph{The maximal speed of
dynamical evolution}, Physica \textbf{D120}, 188 (1998).

\bibitem {deffner17}S. Deffner and S. Campbell,\emph{ Quantum speed limits:
From Heisenberg's uncertainty principle to optimal quantum control}, J. Phys.
A: Math. Theor. \textbf{50}, 453001 (2017).

\bibitem {xu20}T.-N. Xu, J. Li, T. Busch, X. Chen, and T. Fogarty,
\emph{Effects of coherence on quantum speed limits and shortcuts to
adiabaticity in many-particle systems}, Phys. Rev. Research \textbf{2}, 023125 (2020).

\bibitem {tan18}K. C. Tan, S. Choi, H. Kwon, and H. Jeong, \emph{Coherence,
quantum Fisher information, superradiance, and entanglement as
interconvertible resources}, Phys. Rev. \textbf{A97}, 052304 (2018).

\bibitem {rossatto20}D. Z. Rossatto, D. P. Pires, F. M. de Paula, and O. P. de
Sa Neto, \emph{Quantum coherence and speed limit in the mean-field Dicke model
of superradiance}, Phys. Rev. \textbf{A102}, 05716 (2020).

\bibitem {plenio14}T. Baumgratz, M. Cramer, and M. B. Plenio,
\emph{Quantifying coherence}, Phys. Rev. Lett. \textbf{113}, 140401 (2014).

\bibitem {plenio17}A. Streltsov, G. Adesso, and M. B. Plenio,
\emph{Colloquium: Quantum coherence as a resource}, Rev. Mod. Phys.
\textbf{89}, 04103 (2017).

\bibitem {pati16}N. Anand and A. K. Pati, \emph{Coherence and entanglement
monogamy in the discrete analogue of analog Grover search}, arXiv:
quant-ph/1611.04542 (2016).

\bibitem {fan17}H.-L. Shi, S.-Y. Liu, X.-H. Wang, W.-L. Yang, Z.-Y. Yang, and
H. Fan, \emph{Coherence depletion in the Grover quantum search algorithm},
Phys. Rev.\textbf{ A95}, 032307 (2017).

\bibitem {jones47}R. C. Jones, \emph{A new calculus for the treatment of
optical systems V. A more general formulation, and description of another
calculus}, J. Opt. Soc. Am. \textbf{37}, 107 (1947).

\bibitem {goldberg20}A. Z. Goldberg, \emph{Quantum theory of polarimetry: From
quantum operations to Mueller matrices}, Phys. Rev. Research \textbf{2,}
023038 (2020).

\bibitem {anderson94}D. G. Anderson and R. Barakat, \emph{Necessary and
sufficient conditions for a Mueller matrix to be derivable from a Jones
matrix}, J. Opt. Soc. Am. \textbf{A11}, 2305 (1994).

\bibitem {lu96}S.-Y. Lu and R. A. Chipman, \emph{Interpretation of Mueller
matrices based on polar decomposition}, J. Opt. Soc. Am. \textbf{A13}, 1106 (1996).

\bibitem {ryder96}L. H. Ryder, \emph{Quantum Field Theory}, Cambridge
University Press (1996).

\bibitem {wigner59}E. P. Wigner, \emph{Group Theory and Its Applications to
Quantum Mechanics}, Academic Press, New York (1959).

\bibitem {cloude86}S. R. Cloude, \emph{Group theory and polarisation algebra},
Optik \textbf{75}, 26 (1986).

\bibitem {qureshi19}T. Qureshi, \emph{Coherence, interference and visibility},
Quanta \textbf{8}, 24 (2019).

\bibitem {wang15}X. Wang, M. Allegra, K. Jacobs, S. Lloyd, C. Lupo, and M.
Mohseni, \emph{Quantum brachistochrone curves as geodesics: Obtaining accurate
minimum-time protocols for the control of quantum systems}, Phys. Rev. Lett.
\textbf{114}, 170501 (2015).
\end{thebibliography}
\end{document}